%% file: main.tex
\documentclass{article}
\usepackage{iclr2027_conference,times}

\usepackage[utf8]{inputenc}
\usepackage[T1]{fontenc}

\usepackage{graphicx}       
\usepackage{booktabs}       
\usepackage{multirow}       
\usepackage{makecell}       
\usepackage{wrapfig}        
\usepackage{array}          
\usepackage{colortbl}       
\usepackage{xcolor}         
\usepackage{pifont}         
\usepackage{placeins}       
\usepackage{subcaption}     
\usepackage{tikz}           
\usetikzlibrary{positioning,arrows.meta,shapes.geometric,%
                calc,fit,backgrounds,shadows.blur}
\usepackage[most]{tcolorbox} 
\usepackage{listings}        
\tcbuselibrary{listings}     
\input{figures/promptbox}    
\newcommand{\cmark}{\textcolor{teal!70!black}{\ding{51}}}
\newcommand{\xmark}{\textcolor{red!70!black}{\ding{55}}}

\newcommand{\topA}[1]{\cellcolor{orange!65}\textbf{#1}}
\newcommand{\topB}[1]{\cellcolor{orange!42}#1}
\newcommand{\topC}[1]{\cellcolor{orange!22}#1}
\newcommand{\osname}[1]{\multicolumn{1}{>{\cellcolor{cyan!22}}l|}{#1}}

\usepackage{amsmath,amssymb}
\usepackage{bm}
\usepackage{enumitem}\setlist{itemsep=1pt,topsep=2pt,parsep=1pt}

\usepackage[hidelinks,colorlinks=true,
            linkcolor=black,citecolor=teal,urlcolor=teal]{hyperref}
\usepackage{cleveref}

\newif\ifwithappendix
\withappendixtrue

\newcommand{\benchname}{\textsc{CAST-BENCH}}
\newcommand{\Nspec}{900}
\newcommand{\Nl}{9}
\newcommand{\Nmodels}{12}
\newcommand{\NmodelsMulti}{9}
\newcommand{\Nmetrics}{18}

\newcommand{\NperCat}{100}
\newcommand{\Nsingle}{439}
\newcommand{\Nmulti}{461}

\iclrfinalcopy  

\title{CAST-Bench: Tracing Every Sound to Its Source in Multi-Shot Audio-Video Generation}

\author{
  Yucong Liu\textsuperscript{1}\thanks{Equal contribution.} \quad
  Qi Tian\textsuperscript{2$\ast$} \quad
  Shuyuan Tu\textsuperscript{3} \quad
  Kaihang Pan\textsuperscript{2} \quad
  Weijie Kong\textsuperscript{4} \quad
  Chao Pan\textsuperscript{1}\thanks{Corresponding author.} \\
  {\small\textsuperscript{1}Northwestern Polytechnical University,
  \textsuperscript{2}Zhejiang University,
  \textsuperscript{3}Fudan University,
  \textsuperscript{4}Peking University}
}
\date{}

\begin{document}
\maketitle
\lhead{Preprint.}

\begin{abstract}

Existing benchmarks for joint text-to-audio-video (T2AV) generation evaluate perceptual quality and audiovisual alignment, but balanced sound-source sampling and separate reporting by shot structure remain limited. This makes it difficult to determine whether a specific sound is missing, attributed to the wrong source, or lost across shot transitions.
To address this, we introduce \benchname{}, a benchmark designed to
trace every generated sound back to its intended source in multi-shot
audio-video generation. The core idea is to formulate T2AV correctness as
an \emph{audiovisual contract} in which each requested sound must be
emitted by the right source, synchronized with the corresponding visual
action, and carry the right semantic content. 
Concretely, the benchmark organizes 900 prompts into 9 sound-source categories, with 100 prompts per category, spanning speech, animals, instruments, ambient sounds, and other emitter types. The prompts include 439 single-shot and 461 multi-shot narratives, each derived from a structured specification that links every audio event to its emitter, subject, and temporal slot.
The benchmark provides an 18-metric evaluation suite covering three dimensions: Visual Performance, Content Consistency, and Audio \& Semantic Alignment.
We further benchmark 12 representative generators and find that visual quality is nearly saturated across leading systems, yet audio semantic scores still differ markedly.
Across the nine shared models, multi-shot prompts receive lower aggregate scores under the subset-specific scoring rules and yield different model rankings.
Multi-person speech, consistent speaker identity across shots, and cross-shot event retention remain the main unresolved challenges.
All prompts and evaluation code are available at \url{https://github.com/yucong-liu/CAST-Bench-public}.

\end{abstract}

\section{Introduction}
\label{sec:intro}

Text-to-audio-video systems are shifting from silent video synthesis~\citep{openai2024sora} to native joint video-audio generation~\citep{openai2025sora2,deepmind2026veo31,kling2026v3,low2025ovi,davinci2025magihuman,tu2026baton}. 
A prompt can now specify who speaks, the language and emotion they use, whether a dog barks off-screen, or whether a sound persists across a shot. Thus, the output must satisfy an audiovisual contract. The requested event is present, emitted by the intended source, synchronized with the visual action, and semantically consistent with the evolving story.
However, this contract becomes much harder to satisfy in multi-shot narratives, where sounds must not only match their visual sources within each shot but also maintain identity and coherence across shots. A dog's bark may appear in the soundtrack yet be bound to the wrong animal, a line of dialogue may be spoken by the wrong character, or a sound event may drift into a shot where it was never requested. 
These are not corner cases but systematic breakdowns of the source-event-time binding that the contract demands, and no single audio or video quality score can reveal which binding failed or why.

\begin{figure*}[!t]
\centering
\begin{subfigure}[b]{0.48\linewidth}
  \centering
  \includegraphics[width=\linewidth]{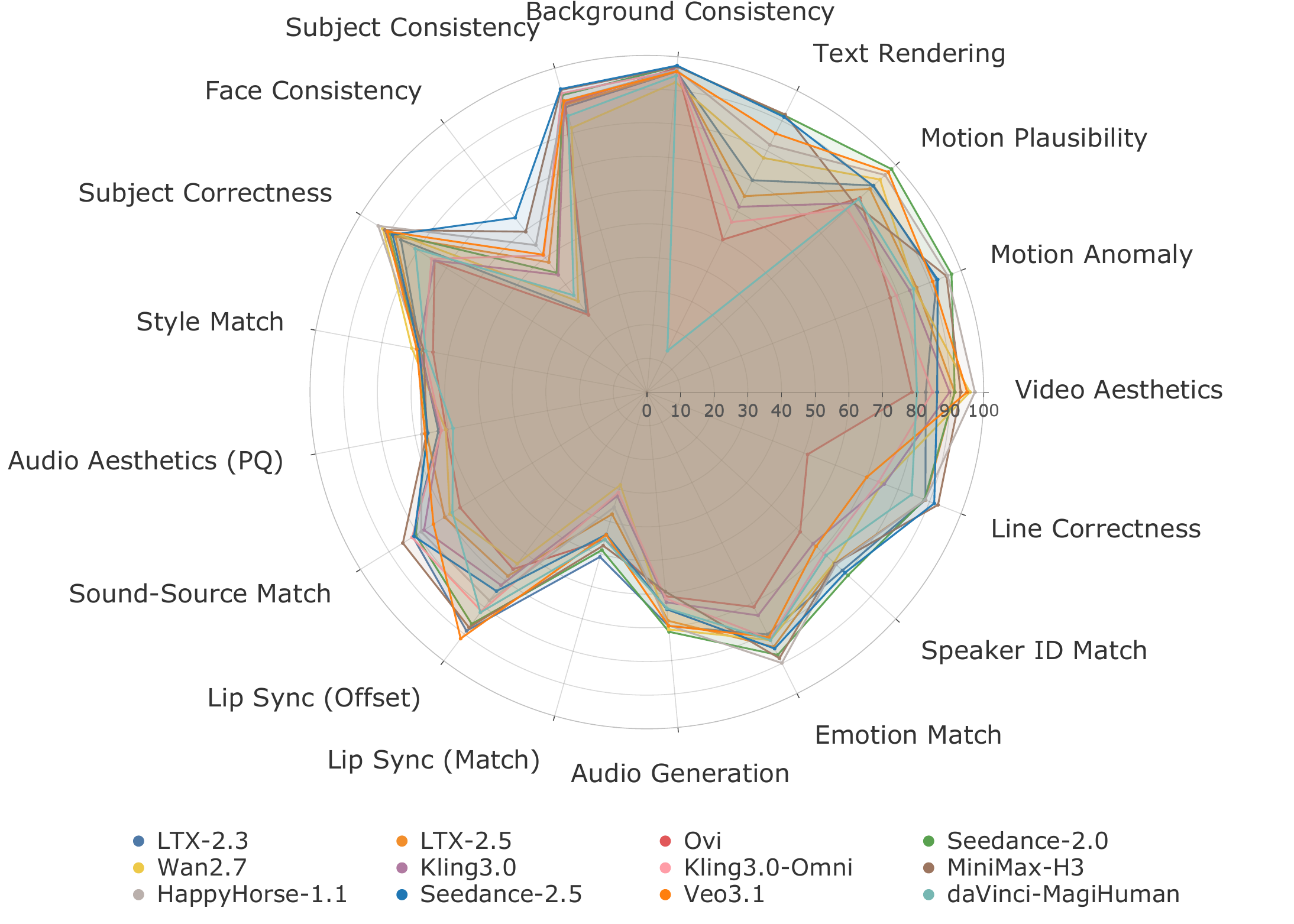}
  \caption{Single-shot subset (\Nmodels{} models).}
  \label{fig:radar-single}
\end{subfigure}\hfill
\begin{subfigure}[b]{0.48\linewidth}
  \centering
  \includegraphics[width=\linewidth]{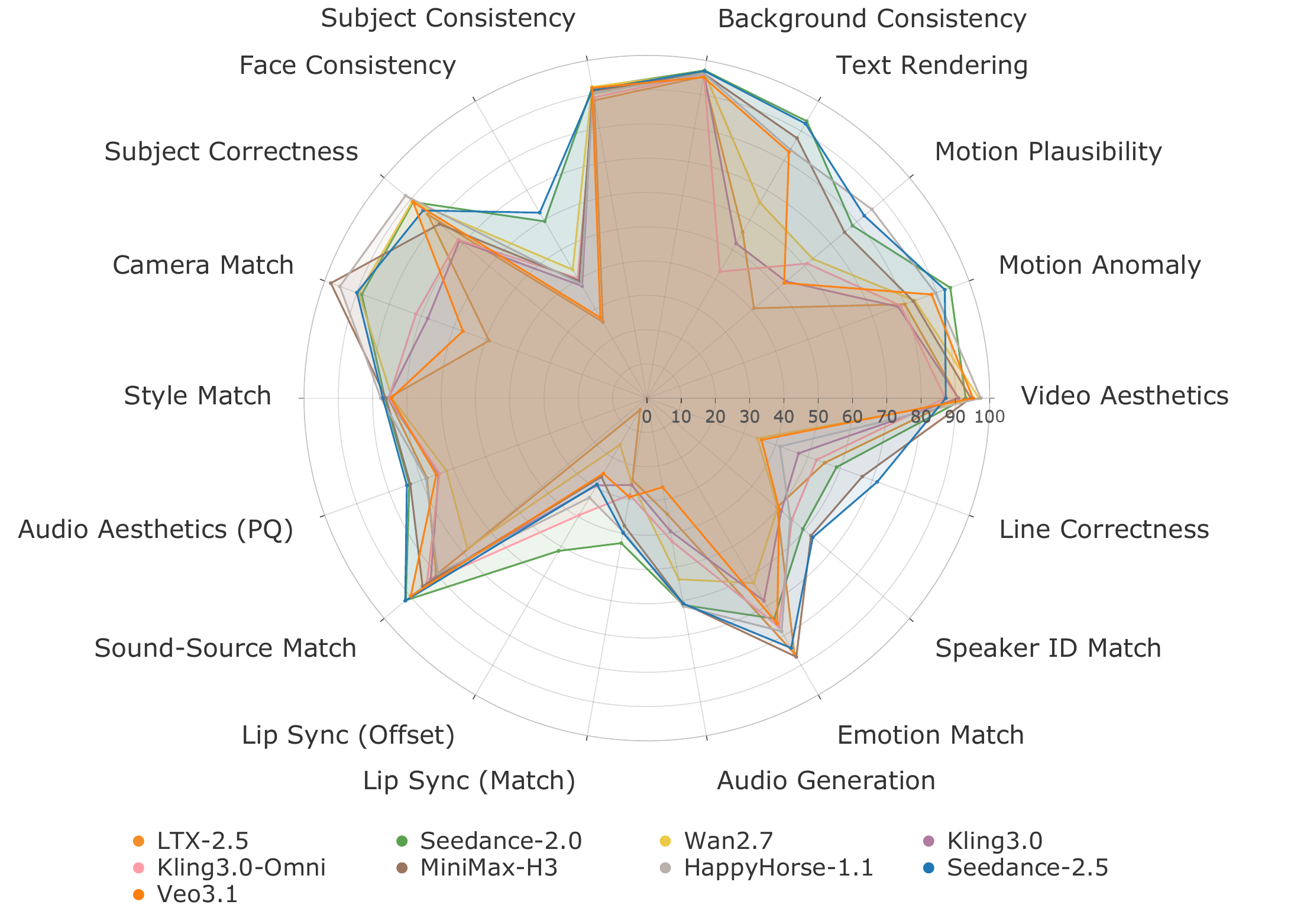}
  \caption{Multi-shot subset (\NmodelsMulti{} models).}
  \label{fig:radar-multi}
\end{subfigure}
\caption{\textbf{Performance radar for video generation models on \benchname{}.}
The radar charts compare \Nmodels{} single-shot and \NmodelsMulti{} multi-shot
video generation models.}
\label{fig:radar}
\end{figure*}


Existing evaluation has not kept pace with this transition. We analyze three gaps in current evaluation suites:
(1) Sound-source coverage is seldom controlled as a balanced sampling axis.
VBench and its extensions organize prompts around visual attributes such as appearance, motion, and intrinsic faithfulness~\citep{huang2024vbench,zheng2024vbenchpp,zheng2025vbench2}, while recent T2AV suites categorize data by application, genre, or broad audio type~\citep{zhou2026avgenbench,wei2026msavbench}. 
As a result, which sound sources are represented, and how frequently, is largely determined by incidental scene content. This makes the coverage of animals, instruments, ambient sounds, and off-screen sources difficult to assess systematically.
(2) Aggregate audio scores can obscure distinct failures. An audio-quality or audio-text alignment score alone cannot distinguish whether an event is generated, whether it comes from the specified sound source, whether its timing is correct, and whether the utterance is produced by the correct speaker. Thus, a low aggregate score provides little insight into which requirement has failed.
(3) Single-shot and multi-shot results share one leaderboard, so one-shot scores can dilute the cross-shot failures where sound-source binding is hardest.

These gaps motivate an evaluation protocol centered on sound sources. Each requested sound should have an explicit emitter, event, and temporal relation that can be traced back to the prompt. The protocol should test both common and challenging combinations of audio and visual conditions, separate single-shot quality from cross-shot continuity, and independently inspect audio evidence when visual cues could provide shortcuts. The goal is to identify which part of the audiovisual contract fails, rather than collapse all failures into a single preference score.
Figure~\ref{fig:radar} shows that mean Sound-Source Match and Style Match improve on multi-shot prompts across the nine shared models. Motion Plausibility has the third-largest normalized mean decrease (24.44 points), after Lip Sync (Offset) and Line Correctness.


We introduce \benchname{}, a sound-source-centered benchmark for diagnosing joint video-audio generation. To give evaluation an explicit reference, each prompt is derived from a structured specification of the requested subjects, sounds, and temporal relations. Complementary metrics distinguish visual quality, content correctness, and audiovisual correspondence, while separate reporting for single-shot and multi-shot prompts exposes differences that a pooled score can obscure. Implementation details are provided in Appendix~\ref{app:metric-procedures}.

Our contributions are threefold.
(1) \textbf{Sound-category coverage:} 900 bilingual prompt--specification pairs cover nine categories and both shot settings.
(2) \textbf{Diagnostic evaluation:} 18 metrics combine specialist estimates with rubric-based judgments of visual and audio-related requirements.
(3) \textbf{Empirical analysis:} experiments compare 12 generators, analyze speaker-count and shot-count groups, and report agreement with human judgments.

\FloatBarrier
\section{Related Work}
\label{sec:related}

\paragraph{Joint video--audio generation.}
Large-scale diffusion models trained with flow-based
objectives~\citep{ho2022videodiffusion,blattmann2023align,peebles2023dit,
lipman2023flowmatching,kong2024hunyuanvideo} underpin many
text-to-video models. Early systems~\citep{openai2024sora,team2025wan,hacohen2024ltxvideo} synthesized silent
frames, requiring a separate stage to add Foley, speech, or music.
Current text-to-audio-video (T2AV) systems~\citep{bytedance2025seedance,
deepmind2026veo31,ai2026wan27,ltx23,low2025ovi,davinci2025magihuman} jointly generate frames and a synchronized
waveform within a shared generation process.
A complementary line of research focuses on how audio is generated and integrated. Video-to-audio models, such as MMAudio, Kling-Foley, and HunyuanVideo-Foley, generate synchronized sound from a fixed visual track~\citep{cheng2025mmaudio,wang2025klingfoley,tencent2025hunyuanfoley}. Joint audio-video systems, including AV-DiT, UniVerse-1, UniForm, MOVA, and JavisDiT, couple the two modalities during generation~\citep{wang2024avdit,wang2025universe1,zhao2025uniform,yu2026mova,liu2025javisdit}. Audio-driven models such as Wan-S2V and StableAvatar instead condition visual generation on supplied audio~\citep{gao2025wans2v,tu2025stableavatar}.
As these models move from single clips toward cinema-like storytelling, maintaining structure and continuity across subjects, scenes, and sounds becomes increasingly important. This makes the evaluation of multi-shot audio-video generation, especially the consistency between what is seen and heard, both increasingly important and challenging.

\input{tables/tab_benchcompare}

\paragraph{T2V and T2AV benchmarks.}

VBench, EvalCrafter, VBench++, VBench-2.0, and T2V-CompBench evaluate visual quality, motion, and text-video alignment, including compositionality and physical plausibility~\citep{huang2024vbench,liu2023evalcrafter,zheng2024vbenchpp,zheng2025vbench2,sun2024t2vcompbench}. AudioCaps and Clotho provide audio-captioning data~\citep{kim2019audiocaps,drossos2020clotho}; AudioTime adds temporal audio-text annotations, and TTA-Bench evaluates text-to-audio systems~\citep{xie2024audiotime,wang2025ttabench}. These resources do not test correspondence to generated visual sources. Recent joint audio-video benchmarks evaluate multi-shot narratives, dialogue, physics, and audiovisual alignment~\citep{wei2026msavbench,zhou2026avgenbench,zhou2026mtavbench,xie2025phyavbench,cao2025t2avcompass,hua2025vabench}. Table~\ref{tab:benchcompare} situates our benchmark among these suites. Our focus is balanced, auditable sampling of emitting sound sources, with separate evaluation of single- and multi-shot outputs.

\FloatBarrier
\section{CAST-BENCH}
\label{sec:bench}

\subsection{Design principles and task formulation}
\label{sec:design}

A generated video may contain a requested sound without placing it in the
intended shot or associating it with the intended subject.  Assessing these
requirements therefore needs an explicit description of the expected events
and their context.  Each \benchname{} item pairs a natural-language prompt
with a structured annotation of the requested subjects, sounds, and shot
requirements.  The generator receives the prompt, while the evaluator also
uses the annotation to identify the requirements being assessed.

To give different sound categories comparable representation, we allocate
100 prompts to each of nine categories.  We also record shot count, visual
style, sounding condition, source complexity, video type, and camera
movement.  These attributes describe the requirements of each prompt and
support analyses across different task conditions.

For multi-shot prompts, the annotation specifies which subjects and audio
events are expected in each shot, including identities that should persist
across cuts.  This distinguishes a requirement to produce an event somewhere
in the video from a requirement to produce it in a particular context.  We
report single-shot and multi-shot results separately, together with
individual metric scores, so that differences between task settings and
failure types remain visible alongside the overall score.

\begin{figure*}[!t]
\centering
\includegraphics[width=\linewidth]{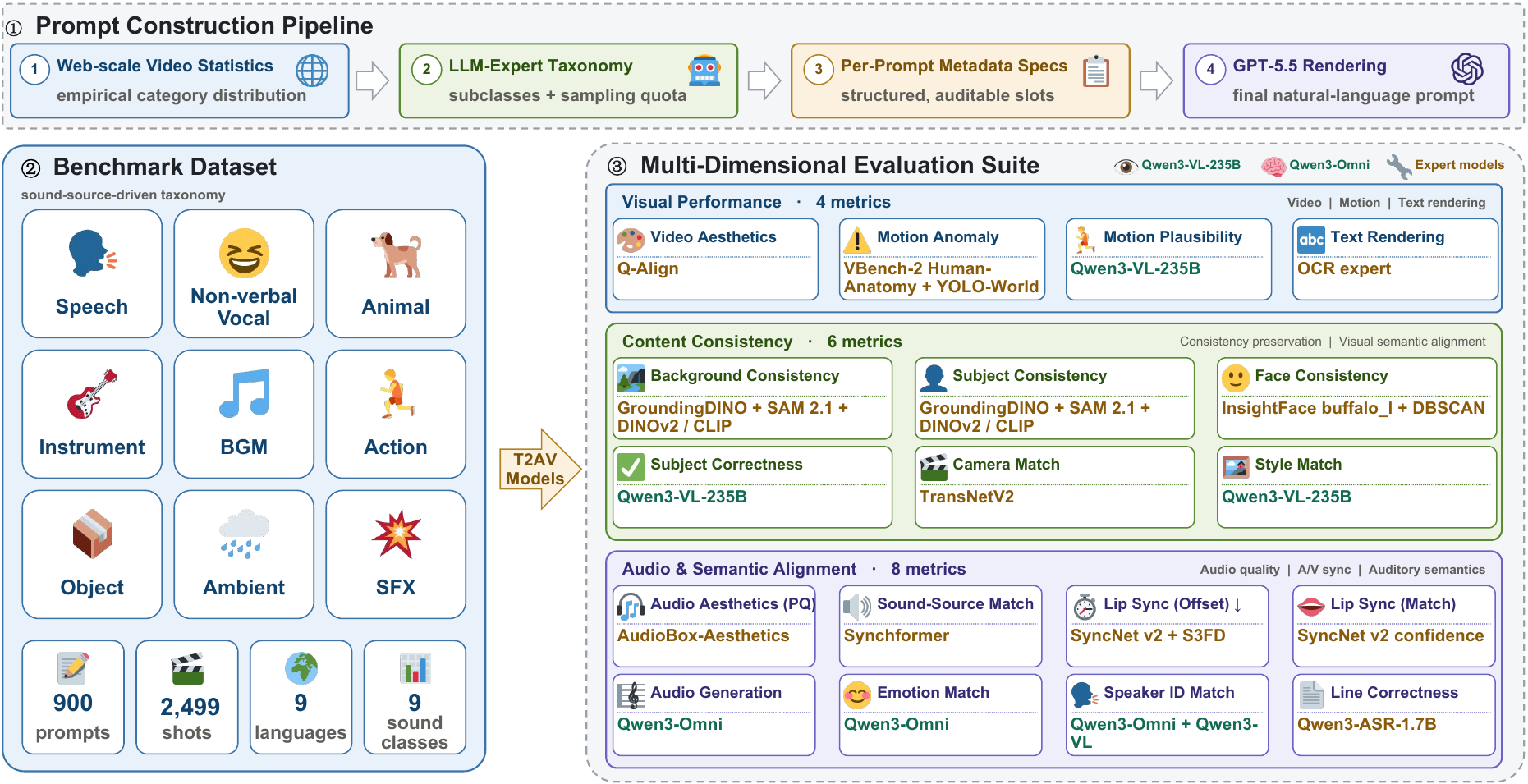}
\caption{\textbf{Overview of the \benchname{} evaluation framework.}
Panel~\raisebox{-0.15ex}{\ding{172}} (Prompt Construction Pipeline) shows four
stages---Web-scale Video Statistics, LLM-Expert Taxonomy, Per-Prompt Metadata
Specs, and GPT-5.5 Rendering---followed by expert validation and coverage
repair. Panel~\raisebox{-0.15ex}{\ding{173}} (Benchmark Dataset) summarizes the
sound-source-driven taxonomy and its 900 prompts, 2,499 shots, nine languages,
and nine sound classes. Panel~\raisebox{-0.15ex}{\ding{174}}
(Multi-Dimensional Evaluation Suite) organizes the shot-aware
\Nmetrics{}-metric harness into Visual Performance, Content Consistency, and
Audio \& Semantic Alignment, with single- and multi-shot scores reported
separately.}
\label{fig:framework}
\end{figure*}

\subsection{Taxonomy and prompt space}
\label{sec:taxonomy}

Free-form prompt writing can leave sound events implicit or describe the same
subject inconsistently across shots.  We therefore construct the task
requirements before expressing them as natural-language prompts.
\Cref{fig:framework} summarizes the construction process: sound categories
guide specification design, and the resulting specifications are converted
into prompts and checked for consistency.  This ordering allows coverage and
evaluation requirements to be inspected before the prompts are used for
generation.

We use observations of publicly available audiovisual examples to inform the
sound categories and event descriptions.  The purpose is to identify recurring
types of audible content that the benchmark should include.
The taxonomy groups prompts by sound category and refines each category into
event or scenario subclasses, with finer labels where applicable.  We combine
these labels with the six task attributes when constructing structured
specifications.  Each specification records the intended subjects and audio
events and associates shot-specific requirements with the relevant shots.
Specifying these elements before writing the prompt makes the intended
coverage explicit and provides a common reference for prompt construction and
evaluation.
We use GPT-5.5 to express each structured specification as a
natural-language prompt.  The conversion instructions require the annotated
subjects, sound events, dialogue, and shot order to be preserved without
introducing additional requirements.  Explicit shot markers identify the
events assigned to each shot in a multi-shot prompt.  Generators receive this
natural-language description, while evaluators also have access to its paired
annotation.  Appendix~\ref{app:prompts:compile} provides the conversion
template and formatting rules.

A well-formed specification does not guarantee that its natural-language prompt
preserves every requirement.  We therefore check structural validity separately
from prompt--specification consistency.  Structural checks cover category
labels, entity references, and shot sequences; semantic checks compare the
rendered prompt with the specification to identify omitted or added events.
We also check for near-duplicate prompts and inspect category coverage.
Section~\ref{sec:stats} summarizes the resulting prompt distribution.

\begin{figure*}[!t]
\centering
\includegraphics[width=\linewidth]{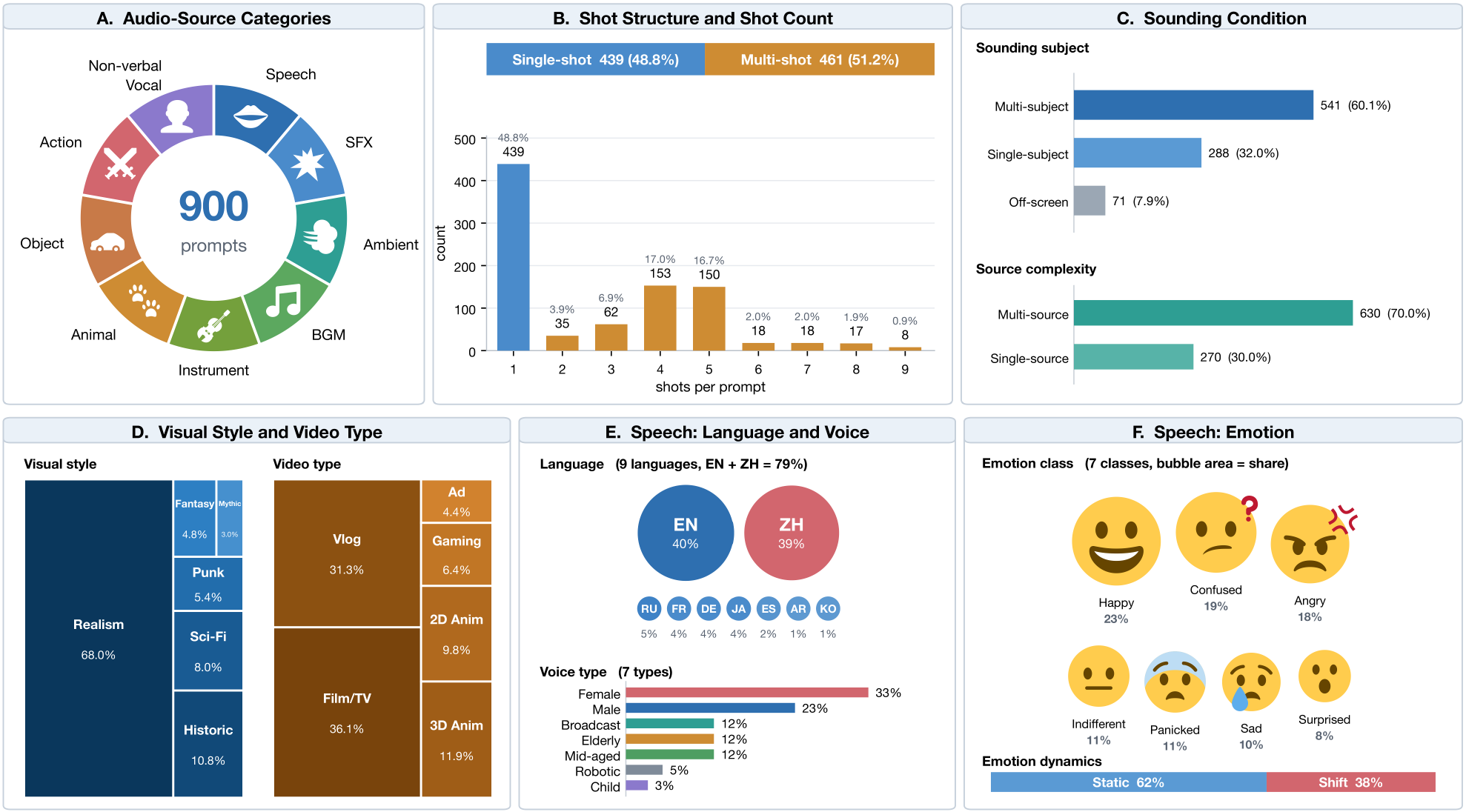}
\captionsetup{belowskip=\baselineskip}
\caption{\textbf{Composition of the \benchname{} prompt suite.}
The suite spans nine audio-source categories (A), single- and multi-shot
structures with one to nine shots (B), sounding-subject and source-complexity
conditions (C), visual styles and video types (D), speech languages and voice
types (E), and emotion classes and dynamics (F). Detailed statistics and
subclass breakdowns are provided in Appendix~\ref{app:prompts:subclasses}.}
\label{fig:promptsuite}
\end{figure*}

\subsection{Dataset statistics and coverage}
\label{sec:stats}

\Cref{fig:promptsuite} summarizes the 900-prompt suite, with 100 prompts in
each sound category.  It contains 439 single-shot prompts and 461 multi-shot
prompts, with the latter requesting between two and nine shots.  The
requested shot counts sum to 2{,}499.

The suite includes 288 prompts with one visible sounding subject, 541 with
multiple sounding subjects, and 71 with an off-screen emitter.  A separate
source-complexity attribute identifies 630 multi-source and 270 single-source
prompts.  The first attribute describes the sounding-subject condition,
whereas the second describes the number of sound sources specified in the
prompt.  These conditions provide different tests of audio-content and
source-correspondence requirements.

The prompts also vary in visual style, video type, speech language, and
emotion requirements.  Within the 100-prompt speech category, nine languages
are represented, with English and Chinese together accounting for 79 prompts.
Thirty-eight prompts in this category specify an emotion change, and 168
prompts across the full suite require on-screen text.  These distributions
describe the coverage of the benchmark rather than equally represented
language or emotion conditions.  Appendix~\ref{app:prompts:subclasses} reports
the detailed subclass counts and their distribution.

\subsection{Evaluation framework}
\label{sec:eval}

\subsubsection{Hierarchical evaluation metrics}
\label{sec:eval:metrics}

Perceptual quality, visual consistency, and audio-related correctness
describe different aspects of a generated video.  A visually convincing
output can still omit a requested sound, render the wrong dialogue, or
produce mismatched mouth motion.  We therefore report 18 metrics in three
groups: Visual Performance, Content Consistency, and Audio \& Semantic
Alignment.  The individual scores help identify which aspects contribute to
an overall performance difference.

\textbf{Visual Performance.} These metrics provide a reference for interpreting
audio-related performance.  \emph{Video Aesthetics}~\citep{wu2023qalign}
measures perceptual visual quality, while \emph{Motion Anomaly}~\citep{zheng2025vbench2} assesses detected
abnormalities in human bodies, faces, and hands.  \emph{Motion Plausibility}
uses five visual questions concerning naturalness, physical plausibility,
temporal coherence, action completeness, and subject integrity.  \emph{Text
Rendering} compares recognized text with the strings requested in the prompt.
These measurements address specific aspects of visual output.

\textbf{Content Consistency.} These metrics cover both appearance stability
and agreement with visual requirements.  \emph{Background Consistency} and
\emph{Subject Consistency} measure feature stability within and across
detected shots, using separate background and subject regions.  \emph{Face
Consistency}~\citep{deng2019arcface} combines agreement with the expected
identity count and the stability of detected face identities.  The remaining
metrics in this group assess specific visual requirements in the prompt.
\emph{Subject Correctness} checks whether the described subjects appear with
the specified identity attributes.  \emph{Camera Match} evaluates whether the
generated video has the requested shot count.  \emph{Style
Match} assesses agreement with the requested visual style and production type.

\begin{figure}[t]
\centering
\includegraphics[width=\linewidth]{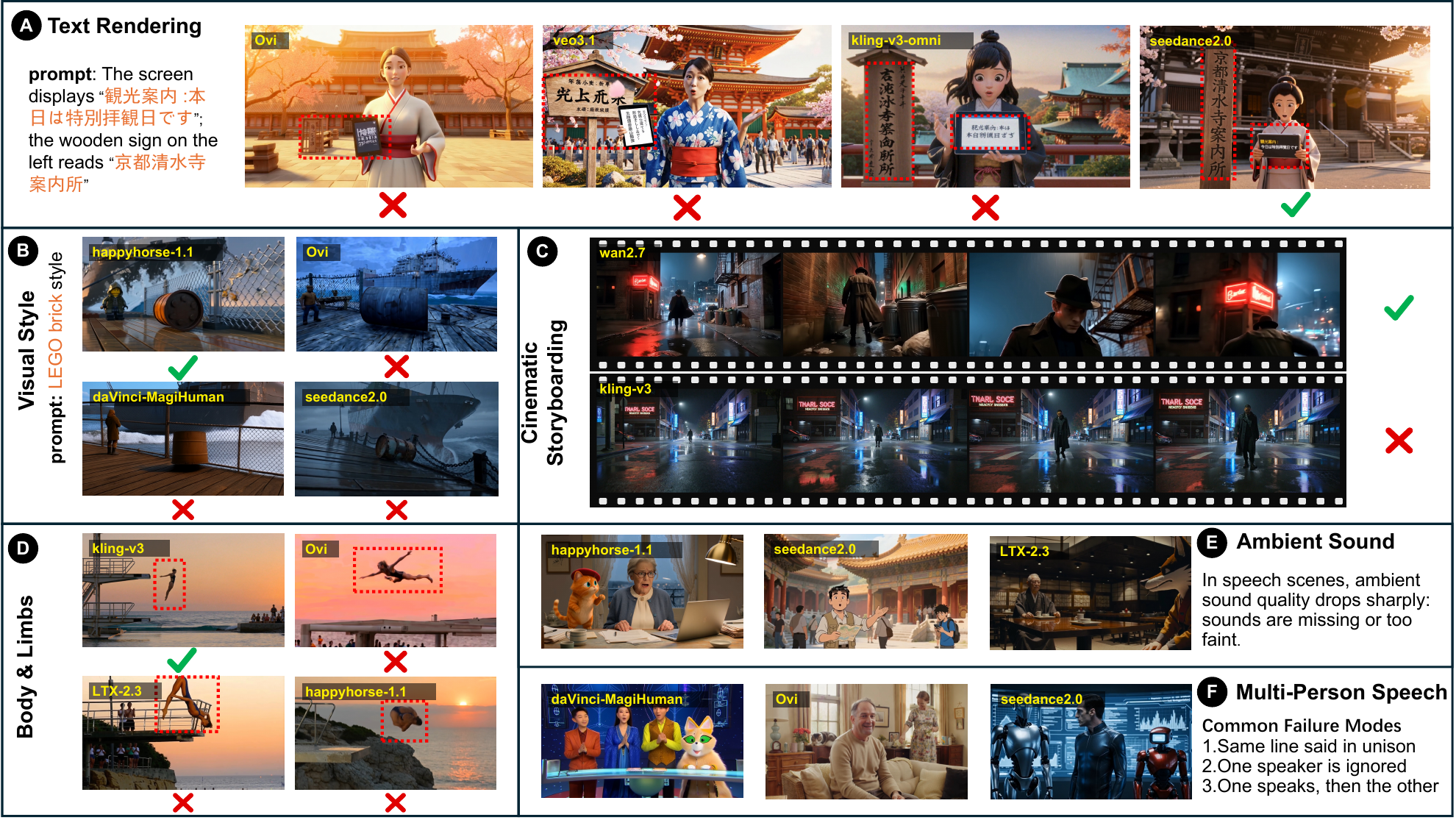}
\caption{\textbf{Representative failure cases on \benchname{}.}
The examples expose six recurring errors: (A) missing or incorrect in-scene
text, (B) style drift, (C) incoherent shot storyboarding, (D) distorted body
and limb motion, (E) missing ambient sound during speech, and (F) failure to
render concurrent multi-person speech.  \cmark{} denotes a correct result,
\xmark{} a failure, and red boxes mark the evidence inspected by the
corresponding metric.}
\label{fig:failure}
\end{figure}

\textbf{Audio \& Semantic Alignment.} Audio quality and audio-content
correctness require different evidence.  A soundtrack can be perceptually
clear while omitting an event requested by the prompt.  We use \emph{Audio Aesthetics (PQ)}, the Production Quality score from
AudioBox Aesthetics~\citep{tjandra2025audiobox}, to assess audio quality,
and \emph{Audio Generation} to assess requested content through
event-presence questions and an overall audio--prompt alignment judgment.  Reporting both scores distinguishes perceptual quality
from whether the expected sounds are actually produced.
Temporal correspondence is evaluated separately from audio-content presence.
\emph{Sound-Source Match}~\citep{iashin2024syncformer} estimates the temporal offset between audio and
corresponding visual activity; a smaller offset indicates better synchrony
but does not verify the source assignment of every sound.  For visible
speech, \emph{Lip Sync (Offset)}~\citep{chung2016syncnet}
measures only the temporal delay between mouth motion and speech, whereas
\emph{Lip Sync (Match)} measures whether the observed mouth shapes and
speech content exhibit the expected phonetic correspondence.  These metrics
assess temporal correspondence under their respective applicability
conditions and do not replace dialogue-content or speaker-assignment
evaluation.
For speech, we distinguish emotional expression, speaker assignment, and
dialogue accuracy.  \emph{Emotion Match} evaluates the specified vocal or
facial emotion requirements.  \emph{Speaker ID Match} compares a
speaker-aware transcript with the requested speakers and utterances, while
\emph{Line Correctness}~\citep{shi2026qwen3asr} measures transcript
agreement with the requested dialogue using word error rate.  Reporting
these scores separately helps identify different speech-related failures.

\input{tables/tab_leaderboard}

\subsubsection{Comprehensive evaluation pipeline}
\label{sec:eval:pipeline}

Prompt compliance involves both measurable signals and semantic requirements. We use specialist estimators for signals such as perceptual quality and temporal offset, and rubric-based judges for requirements such as event presence and speaker assignment. For multi-shot outputs, TransNetV2~\citep{soucek2024transnetv2} detects shot boundaries so that applicable metrics can operate on individual segments.
The specialist branch runs dedicated estimators for visual and audio
aesthetics, human anatomy, OCR, segmentation~\citep{ravi2024sam2} and
feature tracking, face identity, lip synchrony, source/onset displacement,
and ASR. 
The LLM branch contains six rubric-based modules: Motion Plausibility,
Subject Correctness, Style Match, Audio Generation, Emotion Match, and
Speaker ID Match. GPT-5.5 first generates a saturated pool of questions
from the prompt/specification and the metric-specific evidence schema. We
filter this pool for entailment, observability, non-redundancy, and coverage
of all required fields to obtain the final standard question bank.
Qwen3-Omni serves as the audio-understanding
judge~\citep{team2025qwen3omni} and Qwen3-VL as the video-understanding
judge~\citep{team2025qwen3vl}; they answer binary questions or fixed
metric-specific scales.  Speaker ID Match is
a staged exception: Omni first produces turn-level transcripts and speaker
identities through multi-turn audio inspection, and VL then compares those
identities and lines with the cast table and dialogue in the prompt.
Detailed results retain the reporting scale specified for each metric; the two offset metrics are reversed during normalization. Appendix~\ref{app:metric-procedures} provides the metric-level scoring rules.

\FloatBarrier
\section{Experiments}
\label{sec:experiments}

\subsection{Experimental setup}
\label{sec:exp:setup}

We evaluate twelve representative models: seven closed-source generators and
five open-source models (MiniMax-H3, LTX-2.3, LTX-2.5, daVinci-MagiHuman,
and Ovi).  All models are evaluated on the original English prompts, except
the Kling models, which use Chinese prompts because of their input-length
limits.  Outputs are 720p for every model except daVinci-MagiHuman, which is
evaluated at its native 540p, and nominally 10\,s (8\,s for Veo3.1), with three
independent seeds or repeated API calls.  Every metric is first mapped to a
$[0,100]$ higher-is-better scale, reversing the two native lower-is-better
measures; Speaker ID Match is defined on 2--10 and therefore spans 20--100.
The overall score is then a weighted average of
the per-metric means under a \emph{fixed} weight vector that depends only on
the shot setting (Appendix~\ref{app:aggregation}).  Five priority metrics
carry 10\% each and the remaining applicable metrics share the other 50\%
equally.  These weights are not renormalized per sample: each metric enters
with the same weight for every sample of a given shot setting, while its value
is the mean over the samples for which that metric is defined.  Because Ovi, LTX-2.3, and daVinci-MagiHuman
do not support multi-shot generation, they are excluded from the multi-shot
evaluation; the remaining nine models are evaluated on both single- and
multi-shot prompts.

\subsection{Performance across shot settings and evaluation dimensions}
\label{sec:exp:main}

Table~\ref{tab:score_summary} reports the single-shot and multi-shot results.
The three highest single-shot scores are close: 78.11 for Seedance-2.5,
78.03 for Seedance-2.0, and 77.97 for MiniMax-H3.  The same systems lead the
multi-shot subset with scores of 73.28, 72.15, and 69.45, respectively.
Elsewhere, the ordering differs between subsets: Veo3.1 ranks fifth among
the nine shared models on single-shot prompts and eighth on multi-shot
prompts.

Across the nine shared models, mean single-shot and multi-shot scores are
74.30 and 63.59, respectively. The 10.71-point gap compares different metric
compositions (12 versus 13 non-priority metrics); LipO contributes about
2.17 points under the $100-10x$ normalization
(Appendix~\ref{app:aggregation}). All nine models have lower multi-shot
aggregate scores. For Seedance-2.0, Motion Plausibility decreases from
0.983 to 0.783, Audio Generation from 0.715 to 0.613, and Line Correctness
from 0.884 to 0.589.

Performance differences also depend on the evaluation dimension.  On the
single-shot subset, the mean of the eight normalized Audio \& Semantic
Alignment metrics is 60.3 for Ovi, 76.3 for Seedance-2.0, and 75.2 for
LTX-2.3.  LTX-2.3 therefore obtains a similar audio-related group score to
Seedance-2.0 despite a lower visual-group mean, 84.1 versus 94.6.  This
comparison illustrates that visual and audio-related scores capture
different aspects of performance; high visual scores alone do not establish
audiovisual correctness.

Qualitative inspection identifies six recurring failures:
\begin{enumerate}[label=\arabic*.,leftmargin=1.25em,itemsep=0pt,topsep=1pt,parsep=0pt]
\item Insufficient text rendering: fails under multilingual and artistic
      font conditions.
\item Inaccurate video style: styles are difficult to match correctly across
      shots in multi-style videos.
\item Failures in high-dynamic scenes: cinematic shot sequencing becomes
      chaotic or shot breakdown fails, with large-scale limb deformation.
\item Concurrent dialogue is rarely realized: turns are serialized, one line is
      dropped, or both faces mouth the same words, breaking speaker and
      transcript alignment.
\item Insufficient speech-rate adaptation: long lines are clipped rather than
      accommodated by rate adaptation, while short lines are padded with
      unrequested hallucinated speech.
\item Low-quality ambient sound generation: in scenes with speech, background
      ambient sound is under-rendered, drowned out by speech, or largely
      missing.
\end{enumerate}
Figure~\ref{fig:failure} shows representative cases and the metric evidence
behind these diagnoses.

\subsection{Detailed analysis}
\label{sec:exp:analysis}

We compare performance across speaker-count and shot-count groups.

\subsubsection{Speaker multiplicity and shot setting}
\label{sec:exp:speaker}

\input{tables/tab_speaker_audio}

Multiple speakers introduce additional requirements for dialogue and
speaker assignment, while multi-shot prompts require these relations to be
maintained across cuts.  To examine performance under these two conditions,
we compare the G3 scores of one-speaker and multi-speaker prompts
separately within each shot setting (Table~\ref{tab:speaker_audio}).
This analysis is restricted to the \texttt{speech} prompt subset.
Averaged equally across the nine models, the single-shot score is 74.79 for
one-speaker prompts and 70.92 for multi-speaker prompts, a difference of
3.87 points.  The corresponding multi-shot scores are 59.04 and 56.16, a
difference of 2.87 points.  The gap between shot settings is larger: 15.75
points for one-speaker prompts and 14.76 points for multi-speaker prompts.

The direction of these differences is consistent across all nine models.
Every model scores lower on multi-speaker prompts than on one-speaker
prompts in both shot settings, and lower on multi-shot prompts than on
single-shot prompts within each speaker-count group. This consistency indicates that multi-speaker dialogue and cross-shot continuity are systematic bottlenecks rather than weaknesses of a particular architecture. Progress on either dimension alone is therefore unlikely to close the overall gap; models must jointly maintain speaker assignment and shot-level coherence.

\subsubsection{Performance across requested shot counts}
\label{sec:exp:shots}

\begin{wrapfigure}[16]{r}{0.45\textwidth}
\centering
\includegraphics[width=\linewidth]{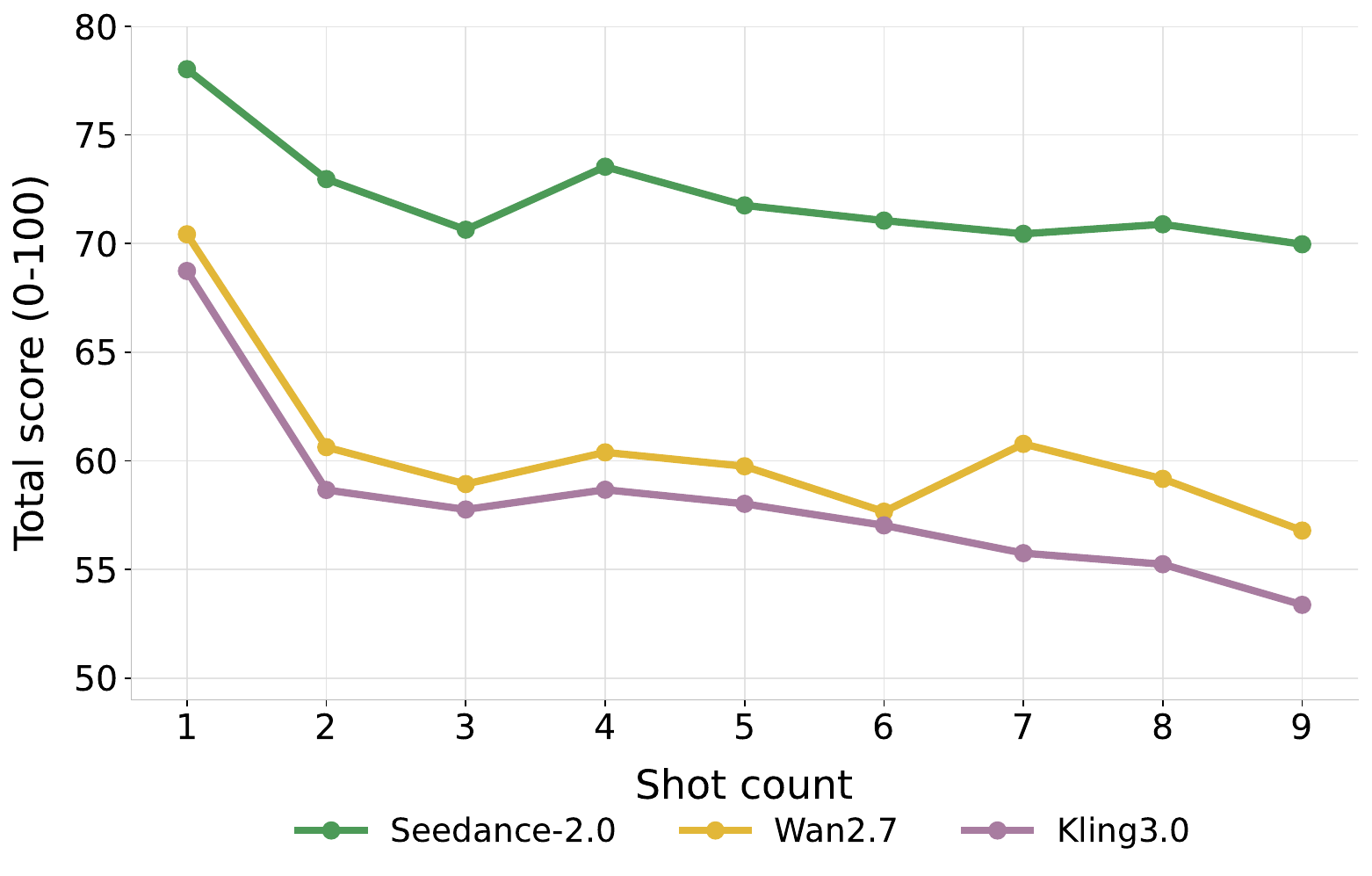}
\captionsetup{width=\linewidth}
\captionof{figure}{Model scores across different shot counts.}
\label{fig:shot_trend}
\end{wrapfigure}

\Cref{fig:shot_trend} compares the plotted scores of Seedance-2.0,
Wan2.7, and Kling3.0 across one to nine requested shots. Seedance-2.0
has the highest score at every shot count, followed by Wan2.7 and Kling3.0.
All three curves show their largest adjacent decrease between one and two
shots, followed by smaller, non-monotonic fluctuations within the
multi-shot range. From one to nine shots, the scores decrease from
78.03 to 69.97 for Seedance-2.0, from 70.43 to 56.79 for Wan2.7, and from
68.74 to 53.37 for Kling3.0, corresponding to differences of 8.06, 13.64,
and 15.37 points. Taken together, the curves highlight a substantial
single-to-multi-shot score gap and stronger score retention by Seedance-2.0.
This comparison indicates that multi-shot generation is still an area where
current audio-video generation models need improvement.

\vspace{0.6\baselineskip}
\subsection{Sampling sensitivity and agreement with human judgments}
\label{sec:exp:reliability}

\begin{wrapfigure}[16]{r}{0.44\textwidth}
\vspace{-0.18\baselineskip}%
\setlength{\columnsep}{12pt}
\centering
\includegraphics[width=\linewidth]{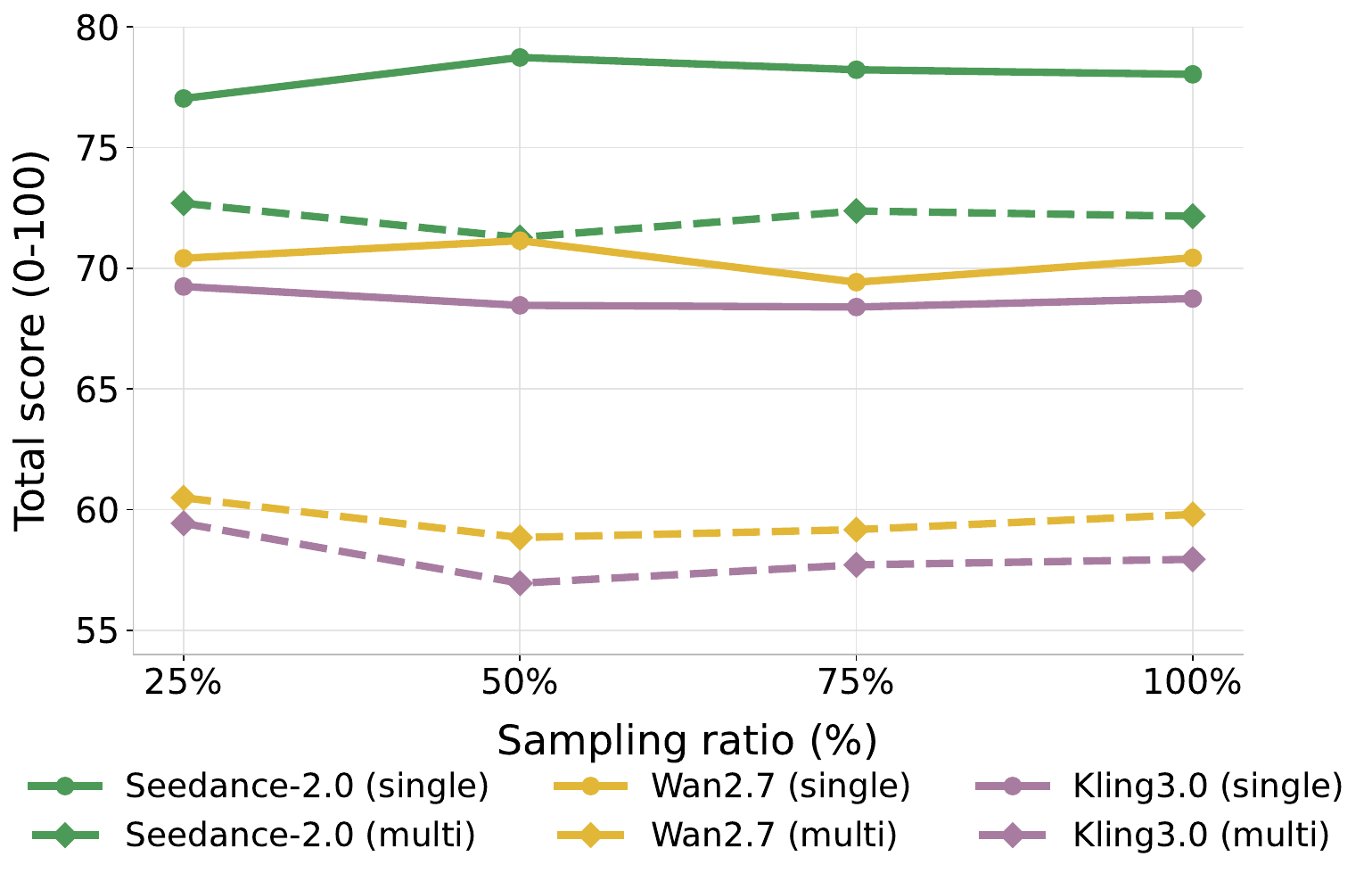}
\captionsetup{width=\linewidth,belowskip=\baselineskip}
\captionof{figure}{Overall score for three representative models as the
evaluated prompt subset grows from 25\% to the full suite.}
\label{fig:sampling_trend}
\end{wrapfigure}

An evaluation should not depend excessively on the particular prompt subset
used to compare models.  \Cref{fig:sampling_trend} examines this sensitivity
at sampling ratios of 25\%, 50\%, 75\%, and 100\% for Seedance-2.0,
Wan2.7, and Kling3.0, separately for single-shot and multi-shot prompts.
Across the displayed subsets, every curve varies by less than three points,
and the ordering of the three systems remains unchanged within each shot
setting.  For each model, the multi-shot score is also lower than the
single-shot score at every sampling ratio.  These observations indicate good
generalizability and a relatively uniform distribution of the prompt set,
and that the overall evaluation system is relatively stable.

The human evaluation yields video-level Spearman point estimates of
0.833 and 0.784 for the single-shot and multi-shot subsets, respectively,
and 0.872, 0.767, and 0.602 for Speaker ID Match, Sound-Source Match,
and Motion Plausibility. Three annotators independently score each video;
Appendix~\ref{app:human-consistency} describes the scoring procedure,
video-level aggregation, correlation analysis, and scale-specific reliability.

\FloatBarrier   

\FloatBarrier
\section{Conclusion}
\label{sec:conclusion}

We present \benchname{}, a 900-prompt benchmark organized by sound category with structured task annotations and 18 evaluation metrics, covering nine sound-source categories. Experiments on 12 representative generators show that strong visual quality does not guarantee reliable audio semantics or speaker correspondence. The nine shared models obtain lower aggregate scores on multi-shot prompts under the subset-specific scoring rules, with pronounced differences in lip synchronization, line correctness, and motion plausibility; sound-source and style matching improve on average. Further analyses examine performance across shot and speaker configurations. \benchname{} provides an auditable framework for measuring and comparing progress in complex audio-video generation.

{\small
\bibliographystyle{iclr2027_conference}
\bibliography{refs}
}

\ifwithappendix
  \clearpage
  \input{sections/appendix_prompts}

\fi

\end{document}

%% file: figures/promptbox.tex
\definecolor{pbBar}{HTML}{3B3B3F}     
\definecolor{pbBg}{HTML}{F2F2F2}      
\definecolor{pbFrame}{HTML}{D0D0D0}   

\tcbset{
    promptbox base/.style={
        enhanced, breakable,
        width=\linewidth,
        colback=pbBg, colframe=pbFrame,
        boxrule=0.4pt, arc=6pt,
        left=10pt, right=10pt, top=8pt, bottom=8pt,
        colbacktitle=pbBar,
        coltitle=white,
        fonttitle=\bfseries\small,
        title style={left color=pbBar, right color=pbBar}, 
        titlerule=0pt,
        toptitle=3pt, bottomtitle=3pt,
        lefttitle=10pt, righttitle=10pt,
    },
}

\newtcolorbox{promptbox}[1]{%
    promptbox base,
    fontupper=\ttfamily\footnotesize,
    before upper={\parindent0pt\raggedright},
    title={#1},
}

\newtcblisting{promptboxV}[1]{%
    promptbox base,
    listing only,
    title={#1},
    listing options={
        basicstyle=\ttfamily\footnotesize,
        breaklines=true,
        breakatwhitespace=false,
        breakindent=0pt,
        columns=flexible,     
        keepspaces=true,
        showstringspaces=false,
        aboveskip=0pt, belowskip=0pt,
        xleftmargin=0pt,
        xrightmargin=0pt,
        resetmargins=true,
    },
}

%% file: tables/tab_benchcompare.tex
\setlength{\columnsep}{12pt}%
\begin{wraptable}[17]{r}{0.46\textwidth}
\vspace{-0.9\baselineskip}%
\begin{minipage}{\linewidth}
\scriptsize
\setlength{\tabcolsep}{1.3pt}
\renewcommand{\arraystretch}{0.92}
\captionsetup{belowskip=12pt}%
\resizebox{\linewidth}{!}{%
\begin{tabular}{@{}lccccc@{}}
\toprule
\textbf{Benchmark} & \textbf{Task} & \textbf{\#M} & \textbf{\#A} & \textbf{\#P} & \textbf{MS} \\
\midrule
VBench~\citep{huang2024vbench} & T2V & 16 & -- & 946 & \xmark \\
EvalCrafter~\citep{liu2023evalcrafter} & T2V & 17 & -- & 700 & \xmark \\
Video-Bench~\citep{han2025videobench} & T2V & 9 & -- & 419 & \xmark \\
TTA~\citep{wang2025ttabench} & T2A & 13 & 5 & 2{,}999 & \xmark \\
Harmony~\citep{hu2025harmony} & TI2AV & 15 & 3 & 150 & \xmark \\
UniAVGen~\citep{zhang2025uniavgen} & TI2AV & 9 & 1 & 100 & \xmark \\
AVGen~\citep{zhou2026avgenbench} & T2AV & 10 & 3 & 235 & \cmark \\
MSAV~\citep{wei2026msavbench} & T2AV & 20 & 3 & 286 & \cmark \\
\midrule
\textbf{Ours} & \textbf{T2AV} & \textbf{\Nmetrics{}} &
  \textbf{\Nl{}} & \textbf{\Nspec{}} & \textbf{\cmark} \\
\bottomrule
\end{tabular}%
}%
\par\smallskip
\captionof{table}{Compact comparison with representative benchmarks.
\textbf{\#M}: listed metric entries; \textbf{\#A}: audio categories/subsets;
\textbf{\#P}: prompt entries; \textbf{MS}: includes multi-shot prompts.
Counts follow each suite's own grouping.}
\label{tab:benchcompare}%
\end{minipage}%
\end{wraptable}


%% file: tables/tab_leaderboard.tex
\begin{table*}[!t]
\centering
\scriptsize
\setlength{\tabcolsep}{2.4pt}
\renewcommand{\arraystretch}{1.1}
\resizebox{\textwidth}{!}{%
\begin{tabular}{@{}l|cccc|cccccc|cccccccc|c@{}}
\toprule
 & \multicolumn{4}{c|}{\textbf{G1 Visual}}
 & \multicolumn{6}{c|}{\textbf{G2 Consistency}}
 & \multicolumn{8}{c|}{\textbf{G3 Audio \& Alignment}}
 & \multirow{2}{*}{\textbf{Ovr\,$\uparrow$}} \\
\cmidrule(lr){2-5}\cmidrule(lr){6-11}\cmidrule(lr){12-19}
\textbf{Method}
 & V-Aes\,$\uparrow$ & Anat.\,$\uparrow$ & Mot.P\,$\uparrow$ & T-OCR\,$\uparrow$
 & BgC\,$\uparrow$ & SbC\,$\uparrow$ & FcC\,$\uparrow$ & SbCor\,$\uparrow$ & CamM\,$\uparrow$ & StyM\,$\uparrow$
 & A-PQ\,$\uparrow$ & SSM\,$\downarrow$ & LipO\,$\downarrow$ & LipM\,$\uparrow$ & AuG\,$\uparrow$ & Emo\,$\uparrow$ & SpID\,$\uparrow$ & ASR\,$\uparrow$
 & \\
\midrule
\multicolumn{20}{@{}l}{\textit{Single-shot subset (12 models)}} \\
\midrule
Seedance-2.5~\citep{bytedance2025seedance25}
  & 0.862 & 0.926 & 0.909 & \topC{0.913}
  & \topA{0.974} & \topA{0.936} & \topA{64.887} & 0.888 & -- & 0.688
  & 6.954 & 0.187 & 2.593 & 4.394 & 0.648 & 0.851 & \topB{7.960} & \topB{0.915}
  & 78.11 \\
Seedance-2.0~\citep{bytedance2025seedance}
  & 0.914 & \topA{0.970} & \topA{0.983} & \topB{0.918}
  & \topB{0.972} & 0.917 & 44.415 & 0.898 & -- & \topC{0.691}
  & 6.973 & 0.193 & 1.370 & \topB{4.878} & \topA{0.715} & \topC{0.872} & \topA{8.080} & 0.884
  & 78.03 \\
\osname{MiniMax-H3~\citep{minimax2026h3}}
  & 0.932 & \topC{0.954} & 0.833 & \topA{0.921}
  & \topC{0.971} & \topB{0.933} & \topB{59.699} & \topC{0.915} & -- & 0.674
  & 6.984 & \topA{0.148} & \topC{1.240} & \topC{4.742} & 0.595 & \topB{0.883} & 7.560 & \topA{0.927}
   & 77.97 \\
HappyHorse-1.1~\citep{happyhorse2026}
  & \topA{0.974} & \topB{0.958} & \topC{0.957} & 0.820
  & 0.961 & 0.899 & \topC{54.709} & \topA{0.938} & -- & 0.687
  & \topA{7.117} & 0.212 & 2.241 & 3.553 & 0.696 & \topA{0.899} & 7.600 & 0.886
   & 77.41 \\
Veo3.1~\citep{deepmind2026veo31}
  & \topC{0.951} & 0.910 & \topB{0.970} & 0.858
  & 0.957 & 0.899 & 51.154 & 0.906 & -- & \topB{0.695}
  & \topB{7.035} & 0.255 & \topA{0.826} & 4.407 & 0.697 & 0.813 & 6.800 & 0.700
   & 74.52 \\
\osname{LTX-2.3~\citep{ltx2026v23}}
  & 0.828 & 0.923 & 0.911 & 0.703
  & 0.956 & 0.880 & 29.841 & 0.859 & -- & 0.683
  & 6.666 & \topC{0.183} & \topB{1.113} & \topA{5.089} & \topC{0.698} & 0.804 & \topC{7.860} & \topC{0.887}
   & 73.66 \\
\osname{LTX-2.5~\citep{ltx25}}
  & 0.914 & 0.859 & 0.896 & 0.650
  & 0.956 & 0.887 & 48.321 & 0.878 & -- & 0.684
  & \topC{6.987} & 0.294 & 3.154 & 3.769 & 0.682 & 0.845 & 7.560 & \topC{0.887}
   & 73.64 \\
Wan2.7~\citep{ai2026wan27}
  & \topB{0.959} & 0.852 & 0.937 & 0.777
  & 0.924 & 0.816 & 33.798 & \topB{0.922} & -- & \topA{0.710}
  & 6.466 & 0.312 & 3.612 & 2.866 & \topB{0.708} & 0.823 & 7.520 & 0.749
   & 70.43 \\
Kling3.0-Omni~\citep{kling2026omni}
  & 0.848 & 0.794 & 0.807 & 0.564
  & 0.960 & \topC{0.923} & 50.831 & 0.752 & -- & 0.674
  & 6.583 & \topB{0.180} & 1.846 & 3.099 & 0.616 & 0.819 & 7.120 & 0.738
   & 69.81 \\
Kling3.0~\citep{kling2026v3}
  & 0.899 & 0.837 & 0.833 & 0.615
  & 0.962 & 0.896 & 43.642 & 0.742 & -- & 0.689
  & 6.610 & 0.222 & 2.815 & 3.209 & 0.627 & 0.741 & 6.680 & 0.756
  & 68.74 \\
\osname{daVinci-MagiHuman~\citep{davinci2025magihuman}}
  & 0.802 & 0.849 & 0.851 & 0.137
  & 0.944 & 0.853 & 36.012 & 0.809 & -- & 0.668
  & 6.269 & 0.321 & 1.800 & 4.553 & 0.645 & 0.823 & 7.200 & 0.843
  & 67.75 \\
\osname{Ovi~\citep{low2025ovi}}
  & 0.787 & 0.775 & 0.856 & 0.506
  & 0.967 & 0.891 & 28.720 & 0.741 & -- & 0.646
  & 6.457 & 0.348 & 3.410 & 4.555 & 0.609 & 0.713 & 6.160 & 0.512
   & 62.34 \\
\midrule
\multicolumn{20}{@{}l}{\textit{Multi-shot subset (9 models)}} \\
\midrule
Seedance-2.5~\citep{bytedance2025seedance25}
  & 0.872 & \topB{0.925} & \topB{0.828} & \topB{0.925}
  & \topB{0.969} & 0.913 & \topA{62.487} & 0.852 & \topC{0.900} & \topB{0.770}
  & \topA{7.704} & \topA{0.080} & 7.092 & \topB{3.994} & \topC{0.609} & \topC{0.842} & \topA{6.320} & \topA{0.715}
   & 73.28 \\
Seedance-2.0~\citep{bytedance2025seedance}
  & 0.930 & \topA{0.942} & \topC{0.783} & \topA{0.933}
  & \topA{0.971} & \topB{0.920} & \topB{59.547} & \topC{0.890} & 0.885 & \topC{0.764}
  & \topB{7.636} & \topB{0.082} & \topA{4.851} & \topA{4.297} & \topB{0.613} & 0.743 & \topC{5.940} & \topC{0.589}
   & 72.15 \\
\osname{MiniMax-H3~\citep{minimax2026h3}}
  & 0.945 & 0.827 & 0.753 & \topC{0.876}
  & 0.961 & 0.910 & 39.547 & 0.790 & \topA{0.981} & 0.761
  & \topC{7.609} & 0.146 & 7.349 & 3.780 & 0.608 & \topA{0.872} & \topB{6.250} & \topB{0.669}
   & 69.45 \\
HappyHorse-1.1~\citep{happyhorse2026}
  & \topA{0.974} & \topC{0.896} & \topA{0.857} & 0.837
  & 0.961 & 0.903 & 38.413 & \topA{0.919} & \topB{0.953} & \topA{0.776}
  & 7.169 & 0.204 & \topC{6.648} & \topC{3.960} & \topA{0.617} & 0.786 & 5.460 & 0.413
   & 66.41 \\
Kling3.0-Omni~\citep{kling2026omni}
  & 0.870 & 0.788 & 0.611 & 0.426
  & 0.960 & 0.891 & 40.388 & 0.719 & 0.717 & 0.753
  & 6.793 & 0.163 & \topB{6.059} & 2.867 & 0.425 & 0.771 & 5.520 & 0.527
   & 60.22 \\
Wan2.7~\citep{ai2026wan27}
  & \topB{0.969} & 0.832 & 0.632 & 0.659
  & \topC{0.965} & \topA{0.922} & \topC{43.254} & \topB{0.893} & 0.894 & 0.743
  & 6.596 & 0.317 & 8.432 & 2.522 & 0.537 & 0.623 & 4.980 & 0.343
  & 59.80 \\
Kling3.0~\citep{kling2026v3}
  & 0.909 & 0.780 & 0.530 & 0.521
  & 0.961 & 0.913 & 37.751 & 0.714 & 0.680 & 0.755
  & 6.819 & 0.177 & 7.058 & 2.574 & 0.394 & 0.683 & 5.120 & 0.471
   & 57.94 \\
Veo3.1~\citep{deepmind2026veo31}
  & \topC{0.951} & 0.884 & 0.523 & 0.829
  & 0.951 & \topC{0.919} & 27.085 & \topC{0.890} & 0.571 & 0.746
  & 6.881 & \topC{0.101} & 7.458 & 2.933 & 0.264 & 0.760 & 5.040 & 0.356
   & 56.91 \\
\osname{LTX-2.5~\citep{ltx25}}
  & 0.908 & 0.801 & 0.408 & 0.559
  & 0.953 & 0.882 & 25.645 & 0.834 & 0.490 & 0.754
  & 7.143 & 0.198 & 9.605 & 2.427 & 0.344 & \topB{0.861} & 4.960 & 0.552
   & 56.13 \\
\bottomrule
\end{tabular}%
}
\caption{\textbf{Leaderboard on \benchname{}.}
\textbf{Visual Performance:} V-Aes: Video Aesthetics, Anat.: Motion
Anomaly, Mot.P: Motion Plausibility, T-OCR: Text Rendering.
\textbf{Content Consistency:} BgC/SbC/FcC: Background/Subject/Face
Consistency, SbCor: Subject Correctness, CamM: Camera Match, StyM: Style Match.
\textbf{Audio \& Semantic Alignment:} A-PQ: Audio Aesthetics (PQ), SSM:
Sound-Source Match, LipO/LipM: Lip Sync (Offset)/(Match), AuG: Audio
Generation, Emo: Emotion Match, SpID: Speaker ID Match, ASR: Line Correctness. Per column,
\topA{deep} / \topB{medium} / \topC{light}
orange highlight the top-1/2/3 scores;
\colorbox{cyan!22}{light-blue model name} = open-source.}
\label{tab:score_summary}
\end{table*}

%% file: tables/tab_speaker_audio.tex
\begin{wraptable}[15]{r}{0.46\textwidth}
\vspace{-0.5\baselineskip}
\setlength{\columnsep}{12pt}
\centering
\scriptsize
\setlength{\tabcolsep}{1.0pt}
\renewcommand{\arraystretch}{0.65}
\captionsetup{width=\linewidth,belowskip=\baselineskip}
\resizebox{\linewidth}{!}{%
\begin{tabular}{@{}>{\raggedright\arraybackslash}p{0.27\linewidth}*{4}{>{\centering\arraybackslash}p{0.16\linewidth}}@{}}
\toprule
 & \multicolumn{4}{c}{\textbf{G3 Audio \& Semantic Alignment}} \\
\cmidrule(lr){2-5}
\textbf{Model} & \multicolumn{2}{c}{\textbf{Single-shot}} &
  \multicolumn{2}{c}{\textbf{Multi-shot}} \\
\cmidrule(lr){2-3}\cmidrule(lr){4-5}
 & \textbf{1 spk.} & \textbf{\raisebox{0.1ex}{\tiny$>$}1 spk.} &
   \textbf{1 spk.} & \textbf{\raisebox{0.1ex}{\tiny$>$}1 spk.} \\
\midrule
\textbf{Seedance-2.0} & \textbf{79.40} & \textbf{79.06} & \textbf{66.62} & \textbf{63.79} \\
MiniMax-H3       & 80.19 & 77.78 & 67.03 & 63.76 \\
Veo3.1            & 76.23 & 70.55 & 54.61 & 50.39 \\
HappyHorse-1.1   & 77.06 & 71.89 & 62.45 & 57.67 \\
Seedance-2.5     & 76.66 & 71.73 & 64.80 & 64.63 \\
LTX-2.5          & 75.77 & 68.79 & 53.59 & 50.10 \\
Kling3.0-Omni     & 71.10 & 68.00 & 58.88 & 55.94 \\
Wan2.7            & 70.58 & 65.06 & 49.40 & 46.73 \\
Kling3.0          & 66.10 & 65.43 & 53.95 & 52.46 \\
\bottomrule
\end{tabular}%
}%
\par\smallskip
\captionof{table}{G3 Audio \& Semantic Alignment score (0--100) by
speaker count and shot setting.}
\label{tab:speaker_audio}
\end{wraptable}

%% file: sections/appendix_prompts.tex
\appendix
\section{Use of AI Tools}
\label{app:ai-tools}

In this work, we used AI tools for the following tasks: reviewing and
correcting the English grammar of the manuscript; surveying the literature
on audio-video and benchmark evaluation to support the related-work analysis;
typesetting and formatting the paper; checking reference correspondence; and
drafting the Introduction and human-evaluation protocol. All AI-assisted
content has been reviewed and verified by the authors prior to submission.
We take full responsibility for the final content of this work, including
any text, claims, or artifacts produced with the aid of AI tools.

\section{Prompt-Construction Pipeline and Templates}
\label{app:prompts}

\begin{figure}[h]
    \centering
    \resizebox{\linewidth}{!}{\input{figures/fig1_pipeline}}
    \caption{\textbf{Reverse-flow prompt-construction pipeline.}
      Domain experts lock the acoustic taxonomy and the
      per-bucket vocabularies (S1, blue); GPT-5.5 fills
      the free-form slots to produce \Nspec{} structured
      JSON \emph{specs}; GPT-5.5 then renders each spec into
      a single natural-language prompt with a tagged
      multi-shot format (S2, teal); the final bilingual
      prompt set (S3, orange) will be released publicly.}
    \label{fig:pipeline}
\end{figure}
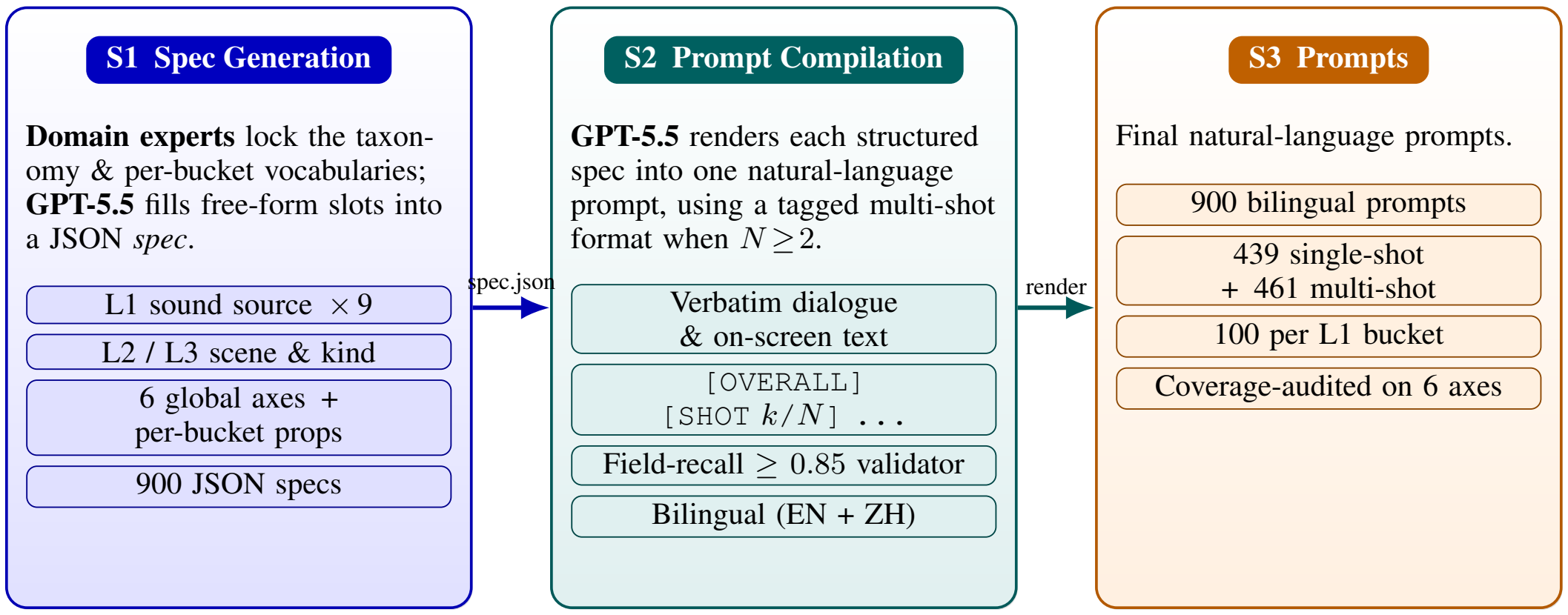

\benchname{}'s prompt suite is produced by a two-stage LLM
pipeline. \Cref{fig:pipeline} shows the overall data flow;
the two system prompts driving Stages~S1 and~S2 are listed
verbatim in \cref{app:prompts:spec,app:prompts:compile}.
Boilerplate rules that are not discussed in the main paper
are elided as ``[\dots]''.

\subsection{S1 --- Spec-Generation System Prompt (GPT-5.5)}
\label{app:prompts:spec}

Given the L1 sound source (fixed by the caller) and the
per-bucket vocabularies, GPT-5.5 emits one
structured JSON \emph{spec} per call. Vocabulary-controlled
slots (\texttt{visual\_style}, \texttt{video\_type},
\texttt{camera\_movement}, \texttt{audio\_situation},
\texttt{sound\_complexity}, \texttt{shot\_count},
\texttt{speech\_attributes.*}, \dots) are enumerated in the
system message; free-form slots
(\texttt{scene\_setting}, \texttt{subjects[*].descriptor},
\texttt{physics\_phenomena}, camera details) are left to the
model.

\begin{promptboxV}{S1: Spec-Generation System Prompt (GPT-5.5)}
You are a T2AV benchmark spec generator. Produce ONE JSON object describing a single video--audio sample. Use ONLY values from the provided vocabularies; do NOT invent enumeration values.

[Constraints]
 (1) `category` and its L1 bucket are fixed by the caller.
 (2) `shot_count` in {single_shot, multi_shot}. If multi_shot, `shots[]` MUST contain EXACTLY N items, where N is provided by the caller and 2 <= N <= 9.
 (3) The 6 global axes (`visual_style`, `video_type`, `camera_movement`, `audio_situation`, `sound_complexity`, `shot_count`) MUST equal the pre-allocated values.
 (4) When `audio_situation` == "off_screen", the sound emitter MUST NOT be visible in the frame.
 (5) NARRATIVE CONTINUITY (HARD, multi-shot): all shots MUST tell ONE coherent story; each shot picks up from the previous one.
 (6) Every `subjects[*].descriptor` MUST be >= 8 words with concrete visual attributes (age, clothing, material, lighting).
 (7) For speech buckets, `subjects[*].speech` or `shots[*].dialogue[*].text` MUST be written in the script/language specified by `speech_attributes.language`.
 [ ... boilerplate about JSON escaping / vocab lists ... ]

[Vocabulary]
 [[VOCAB_ENUMERATION]]        # injected at runtime

[Output] JSON only. No markdown, no commentary, no ```json fences.

Now produce the spec for:
 category = [[CATEGORY]]
 shot_count = [[SHOT_COUNT]]     (N = [[N_SHOTS]] if multi)
 pre-allocated axes = [[AXES]]
\end{promptboxV}

\subsection{S2 --- Prompt-Compiler System Prompt (GPT-5.5)}
\label{app:prompts:compile}

Each JSON spec is then handed to GPT-5.5, which
\emph{compiles} it into one natural-language prompt.
Multi-shot specs use the tagged format
\texttt{[OVERALL] \dots\ [SHOT 1/$N$] \dots\ [SHOT $N$/$N$]}
so downstream generators can detect shot boundaries with a
simple regex.

\begin{promptboxV}{S2: Prompt-Compiler System Prompt (GPT-5.5)}
You are a T2AV prompt compiler. Convert a structured spec JSON into a single natural-language video-generation prompt.

Rules:
 (1) Embed every concrete element from the spec -- subjects, audio_events, camera, scene_setting, and all top-level global axes (visual_style, video_type, ...).
 (2) NEVER invent details not in the spec. ONLY describe what the spec contains; do NOT add atmospheric flourish, similes, metaphors, emotional commentary, or extra sound effects that the spec did not ask for. Stay strictly factual.
 (3) Output ONLY the prompt text; no preamble, no JSON.
 (4) sound_complexity == single_source: the prompt MUST NOT mention any sound source other than the primary audio subject.
 (5) audio_situation == off_screen: do NOT describe the visual appearance of the off-screen emitter; describe only its audible qualities.
 (6) DIALOGUE INTEGRITY (HARD): quote every line from `subjects[*].speech` / `shots[*].dialogue` VERBATIM, in the original script; never paraphrase, translate, abridge, or write "...". Every quoted line MUST end with terminal punctuation.
 (7) ON-SCREEN TEXT: entries in `text_on_screen` MUST be rendered as visible text in the scene, never as spoken dialogue. Dialogue and on-screen text are two disjoint sets.
 (8) EMOTION ARC (HARD, speech): when `speech_attributes.emotion_change == true`, the prompt MUST describe a single emotional turn from `emotion_from` to `emotion_to`; the OVERALL header MUST mention this arc in plain words.
 (9) Multi-shot format:
       [OVERALL] <one short sentence embedding all global axes>
       [SHOT 1/N, t1s, <camera>] <shot description> ...
       [SHOT N/N, tN s, <camera>] <shot description> Output one continuous paragraph -- no bullet points.
 [ ... boilerplate about length limits / language ... ]

Now compile the following spec:
 [[SPEC_JSON]]
\end{promptboxV}

\subsection{Prompt-subclass statistics}
\label{app:prompts:subclasses}

This section gives the fine-grained subclass statistics omitted from the
main text. Percentages are rounded to one decimal place. For L2, the
denominator is the number of valid L2 event occurrences in the corresponding
L1 bucket, so a prompt containing several audio events contributes to several
L2 counts. For L3, we count only the first valid event of the corresponding
L1 bucket (the designated main event), giving a denominator of 100 prompts
per bucket. Categories without an L3 vocabulary use their L2 subclasses as
the finest event-level partition.

\paragraph{Global-axis marginals.}
The suite mixes 439 single-shot prompts (48.8\%) with 461 multi-shot prompts
(51.2\%), the latter spanning two to nine ordered shots; four-shot (153,
17.0\%) and five-shot (150, 16.7\%) narratives are the largest groups,
followed by three-shot (62, 6.9\%), two-shot (35, 3.9\%), six- and
seven-shot (18 each, 2.0\%), eight-shot (17, 1.9\%), and nine-shot (8,
0.9\%) cases. On source binding, 541 prompts (60.1\%) contain multiple
sounding subjects, 288 (32.0\%) contain one visible sounding subject, and 71
(7.9\%) use an off-screen emitter; independently, 630 prompts (70.0\%)
contain multiple sound sources and 270 (30.0\%) a single source. Visual style
is realistic in 612 prompts (68.0\%), historical 97 (10.8\%), sci-fi 72
(8.0\%), punk 49 (5.4\%), fantasy 43 (4.8\%), and xuanhuan 27 (3.0\%). Video
type is film/TV 325 (36.1\%), vlog 282 (31.3\%), 3D animation 107 (11.9\%),
2D animation 88 (9.8\%), games 58 (6.4\%), and advertisements 40 (4.4\%).
Speech spans nine languages---English 40 (40.0\%), Chinese 39 (39.0\%),
Russian 5 (5.0\%), French, German, and Japanese 4 each (4.0\%), Spanish 2
(2.0\%), and Arabic and Korean 1 each (1.0\%). Emotion classes are joyful 23
(23.0\%), confused 19 (19.0\%), angry 18 (18.0\%), neutral and panicked 11
each (11.0\%), sad 10 (10.0\%), and surprised 8 (8.0\%); 38 prompts (38.0\%)
additionally specify an emotion change, and 168 prompts (18.7\%) require
on-screen text.

\paragraph{Speech.}
L2 occurrences (184): daily dialogue 93 (50.5\%), broadcast 43 (23.4\%),
commentary 29 (15.8\%), voiceover 18 (9.8\%), and singing 1 (0.5\%). The
main-event L3 distribution (100): film commentary 8 (8.0\%), press conference
8 (8.0\%), radio broadcast 8 (8.0\%), tour guide 8 (8.0\%), casual conversation
7 (7.0\%), documentary narration 7 (7.0\%), film voiceover 7 (7.0\%),
interview 7 (7.0\%), lecture 7 (7.0\%), meeting 7 (7.0\%), platform livestream
7 (7.0\%), quarrel 7 (7.0\%), news broadcast 6 (6.0\%), and sports commentary
6 (6.0\%). Voice timbre is female 33 (33.0\%), male 23 (23.0\%), broadcast
12 (12.0\%), elderly 12 (12.0\%), middle-aged 12 (12.0\%), robotic 5 (5.0\%),
and child 3 (3.0\%). Languages are English 40 (40.0\%), Chinese 39 (39.0\%),
Russian 5 (5.0\%), French 4 (4.0\%), German 4 (4.0\%), Japanese 4 (4.0\%),
Spanish 2 (2.0\%), Arabic 1 (1.0\%), and Korean 1 (1.0\%). Emotions are joyful
23 (23.0\%), confused 19 (19.0\%), angry 18 (18.0\%), neutral 11 (11.0\%),
panicked 11 (11.0\%), sad 10 (10.0\%), and surprised 8 (8.0\%); 62 prompts
(62.0\%) have no marked emotion change and 38 (38.0\%) contain a change.

\paragraph{Non-verbal vocalization.}
L2 occurrences (227): physiological 103 (45.4\%), emotional 96 (42.3\%),
and mimicry 28 (12.3\%). The main-event L3 distribution (100) is cheer,
cough, cry, fart, hiccup, laugh, moan, scream, smirk-laugh, sneeze, sniffle,
snore, tongue click, whistle, and yawn at 6 each (6.0\%), followed by beatbox
and sigh at 5 each (5.0\%).

\paragraph{Animal.}
L2 occurrences (203): bird call 52 (25.6\%), insect sound 45 (22.2\%), wild
beast roar 34 (16.7\%), livestock 29 (14.3\%), pet vocalization 22 (10.8\%),
and amphibian 21 (10.3\%). The main-event L3 distribution (100) is bear,
chicken, dog, elephant, mosquito, pig, and snake at 6 each (6.0\%); cat,
cricket, crow, cuckoo, duck, eagle, frog, horse, tiger, and wolf at 5 each
(5.0\%); and bee and cicada at 4 each (4.0\%).

\paragraph{Object.}
L2 occurrences (215): interaction 104 (48.4\%), transportation 67 (31.2\%),
and weapon 44 (20.5\%). The main-event L3 distribution (100) is car,
collision, dart, motorcycle, and tear at 6 each (6.0\%); airplane, arrow,
bicycle, bomb, friction, knife, liquid, missile, pistol, police car, rifle,
shatter, ship, and train at 5 each (5.0\%).

\paragraph{Action.}
L2 occurrences (176): daily activity 108 (61.4\%) and sports 68 (38.6\%).
The main-event L3 distribution (100) is fancy diving and martial-arts fighting
at 11 each (11.0\%), basketball layup contest at 9 (9.0\%), sword fighting at
8 (8.0\%), wrestling at 7 (7.0\%), unarmed fighting at 5 (5.0\%), boxing,
clapping, knocking, stick fighting, and swimming race at 4 each (4.0\%),
basketball, gymnastics, running, soccer, table tennis, and walking at 3 each
(3.0\%), billiards, chopping vegetables, eating, soccer dribbling, and typing
at 2 each (2.0\%), and parkour at 1 (1.0\%).

\paragraph{Ambient sound.}
There is no separate L3 layer. L2 occurrences (289) are weather 81 (28.0\%),
urban street 74 (25.6\%), water body 45 (15.6\%), crowd/applause 34 (11.8\%),
forest/field 31 (10.7\%), and indoor ambience 24 (8.3\%).

\paragraph{Instrument.}
There is no separate L3 layer. L2 occurrences (157) are piano 71 (45.2\%),
string instrument 56 (35.7\%), percussion 16 (10.2\%), synthesizer 7 (4.5\%),
wind instrument 7 (4.5\%). Among the 98
prompts with an annotated playing mode, 53 are solo (54.1\%) and 45 are
ensemble (45.9\%); two prompts have no mode annotation.

\paragraph{Background music.}
There is no separate L3 layer. L2 occurrences (144) are film score 91 (63.2\%),
vlog BGM 31 (21.5\%), and game music 22 (15.3\%). The music-kind attributes
are instrumental 43 (43.0\%), song 30 (30.0\%), and rhythmic beat 27 (27.0\%).

\paragraph{Sound effect.}
There is no separate L3 layer. L2 occurrences (341) are exaggerated SFX 86
(25.2\%), UI notification 81 (23.8\%), transition SFX 67 (19.6\%), sci-fi SFX
58 (17.0\%), magic-skill SFX 36 (10.6\%), and ambience SFX 13 (3.8\%).

\section{General-Model Evaluation Details}
\label{app:eval}

For every VLM-graded metric we materialise spec-grounded
questions or score prompts at benchmark-build time; at
evaluation time these are handed to the judge model together
with the generated clip. Binary metrics use yes/no answers,
whereas style, overall audio alignment, and speaker identity use explicit
integer scales. This appendix lists, for each metric, one
representative spec together with (i) the exact questions we
ask the judge, and (ii) the corresponding grader-rubric prompt
that defines the judge's evidence and scoring rules. Boilerplate
is elided as ``[\dots]''.

\subsection{Audio Generation}
\label{app:eval:audio-generation}

\emph{Audio Generation} asks whether every audible sound
described in the prompt is actually rendered in the generated
video, one yes/no question per sound source, plus one overall
1--5 alignment question. Questions are written per shot and
must never invent sounds that the prompt does not describe.

\begin{promptboxV}{Prompt \& Questions: Audio Generation}
------ Prompt ------ A gritty urban basketball court at dusk, chain-link fences and graffiti walls surrounding cracked asphalt with faded lines. A male player in a red jersey and black shorts sweats as he drives for a layup, sneakers screeching on pavement, ball bouncing rhythmically before release. A female defender in a blue jersey leaps to block, body tensing mid-air. A male referee in a striped shirt watches intently from the sideline, whistle dangling. The digital scoreboard above the hoop reads HOME 78 - GUEST 76 with 12.0 seconds left; a banner at the bottom displays GAME ON THE LINE. Distant crowd murmurs, a car engine idles faintly, and a siren wails far away. Body
movements produce natural sounds: dribble thumps, sneaker friction, jump impact, net swish. No camera movement -- fixed frame captures the physics of momentum, gravity, collision, and friction in real-time.

------ Questions ------
Q1.  Is there a sound of sneakers screeching on pavement in the video?
Q2.  Is there a sound of a basketball bouncing rhythmically in the video?
Q3.  Is there a sound of a dribble thump in the video?
Q4.  Is there a sound of sneaker friction in the video?
Q5.  Is there a sound of a jump impact in the video?
Q6.  Is there a sound of a net swish in the video?
Q7.  Is there a sound of distant crowd murmurs in the video?
Q8.  Is there a sound of a car engine idling faintly in the video?
Q9.  Is there a sound of a siren wailing far away in the video?
Q10. How well does the audio in the video match the prompt?     [1-5 rubric]
\end{promptboxV}

\begin{promptboxV}{Grader Rubric: Audio Generation}
You are a strict audio judge for one short, single-shot clip. The original
generation prompt is the ground truth. Use the visuals only to identify a
sound source; judge the audible soundtrack itself.

There are two question types. For a specific sound or speaker question,
answer yes only when the requested sound is clearly audible and consistent
with the prompt; otherwise answer no. For a speaker question, also require
the same gender, age, role, language, content, and visible mouth alignment,
and do not merge two prompt-described speakers. Silent or non-speaking clips
are no for speaker questions. Output one object per numbered question:
{"reason":"brief audible evidence","answer":"yes"|"no"}.

The exact overall question is: "How well does the audio in the video match
the prompt?" Score it from 1 to 5 using the following rubric:
  5  Every required sound is present, correctly timed and balanced, with no
     extraneous or distorted audio.
  4  All major sounds are present, with only a minor audible flaw.
  3  Some required sounds are present but coverage or timing is incomplete.
  2  The soundtrack is mostly inconsistent with the prompt.
  1  Audio is wrong, silent, unintelligible, or contradicts the prompt.
Use 5 only for flawless audio; any faint, late, missing, distorted, or extra
sound forces a score of at most 4, and a silent or contradictory track scores
1. For this question output {"reason":"heard audio and score justification",
"score":1--5}. Reasons must describe what is heard rather than restating the
prompt or relying on visual evidence. Return only the required JSON object or
array, with no extra keys or prose.
\end{promptboxV}

\subsection{Emotion Match}
\label{app:eval:emotion}

\emph{Emotion Match} probes whether the vocal emotion
in a speech clip matches the spec, and---when the spec
declares an emotion \emph{change}---whether the on-screen
facial expression follows the same arc. Questions are
emitted at the spec level (not per shot) since emotion
labels are global.

\begin{promptboxV}{Prompt \& Questions: Emotion Match}
------ Prompt ------
[OVERALL] Realistic vlog-style multi-shot sequence capturing a corporate meeting's emotional arc from angry to confused, featuring multi-subject dialogue with broadcast-tone English speech, fixed-frame shots, and a cartoon coffee mug confessing a system error.

[SHOT 1/3, medium_shot] A middle-aged male executive in a dark suit, stern expression, slams his fist while pointing at a laptop screen showing red financial charts. He leans forward, jaw tight, eyes blazing as he delivers, "This quarter's numbers are unacceptable -- we're hemorrhaging revenue and no one has an explanation! I want answers by noon or heads will roll." A young female analyst fidgets with her pen; a male engineer leans against the wall. A cartoon-style animated coffee mug blinks its eyes slowly.
Audio: only the executive's voice, continuous and forceful.

[... SHOT 2/3 and SHOT 3/3 elided;
     emotion arc completes: angry -> confused ...]

------ Questions ------
Q1.  Does the speaking voice in the video start with an angry emotion?
Q2.  Does the speaking voice in the video transition from an angry emotion to a confused emotion?
Q3.  Does the on-screen character's facial expression shift in a way that matches the vocal emotional transition?
\end{promptboxV}

\begin{promptboxV}{Grader Rubric: Emotion Match}
You are a strict audio-visual emotion judge. For every question, return only
{"reason":"concrete evidence in at most 20 words","answer":"yes"|"no"}.

Global override: if no human speech is detectable (silence, music, ambience,
sound effects, non-verbal vocalizations, or speech too faint or unintelligible),
answer no for every question.

Question types are modality-specific:
  A. "Is the speaking voice in the video expressing a <emotion> emotion?"
     Judge the audible speech only. If several people speak, one clearly
     matching speaker is sufficient.
  B. "Do the on-screen characters visually display a <emotion> emotion?"
     Judge frames only; at least one visible person must show the requested
     expression. No person, an unclear face, or the wrong emotion is no.
  C. "Are there no people visible in the video?" Judge frames only; answer
     yes only when no human or humanoid appears anywhere in the clip.

Do not transfer evidence between modalities. Broken or undecodable frames make
the visual questions no. Return one object or an array matching the numbered
questions, with no extra fields or prose.
\end{promptboxV}

\subsection{Speaker ID Match}
\label{app:eval:speaker-id}

\emph{Speaker ID Match} evaluates whether the generated clip preserves
the identity of every speaking subject across turns and shots, and whether
each utterance is assigned to the correct speaker.  The implementation uses
two stages: Qwen3-Omni first produces an ordered speaker-aware transcript,
then a rubric-based judge compares that transcript with the ground
truth prompt and subject table.  The two stages below are faithful English
renderings of the prompts used by the evaluation code; the executable files
will be provided with the benchmark release.

\begin{promptboxV}{Qwen3-Omni System Prompt (Stage 1)}
You will see a video containing people speaking, including both visual frames and audio. Transcribe ALL audible spoken dialogue and narration in temporal order. Annotate every segment with:
 - speaker: a speaker ID (S1, S2, ...)
 - subject: a short description of the speaking subject (appearance or ``off-screen voice'')
 - text: the spoken content of that segment

Speaker-ID rules:
 1. The first subject to speak is S1. Off-screen narration, voiceover, and inner monologue count as speakers; if one speaks first, it is S1.
 2. Assign S2, S3, ... to newly appearing speakers in order of first speech.
 3. When a speaker speaks again, reuse exactly the ID assigned at first appearance.
 4. Off-screen narration is an independent speaker and must not be merged with an on-screen character.
 5. Retain meaningful singing lyrics and meaningful vocalizations (for example, ``ah'', ``um'', or ``hey''). Do not transcribe instrumental background music or pure environmental sound.

Subject-description rules:
 1. For off-screen narration, voiceover, or inner monologue, use one fixed ``off-screen voice'' label.
 2. For visible people, use a brief description containing gender, age range, and salient appearance or clothing; use at most 15 Chinese characters in the executable prompt.
 3. The subject string must remain identical for repeated segments from the same speaker.
 4. The executable prompt uses Chinese for the subject description even when the spoken language is English, French, Japanese, or another language.
 5. If appearance cannot be determined, use a consistent fallback such as ``unknown male voice'', ``unknown female voice'', or ``unknown speaker''.

Output format (strict): output exactly one JSON object, with no Markdown fences, explanation, timestamps, or extra fields:
{"segments":[{"speaker":"S1","subject":"off-screen voice","text":"..."},{"speaker":"S2","subject":"middle-aged man in a white shirt","text":"..."}]}

Additional requirements:
 - speaker must be a string of the form S1, S2, S3, ... .
 - subject must be a non-empty string following the rules above (Chinese in the executable prompt).
 - text must contain only the spoken dialogue; do not emit [Music], [Noise], or [Unintelligible] tags.
 - segments must be ordered by the actual time of occurrence in the video.
 - If no one speaks, output {"segments":[]}.
\end{promptboxV}

\begin{promptboxV}{Grader Rubric: Speaker ID Match}
You are a strict judge for speaker-ID consistency. You will receive two JSON objects:

GT contains video_id, shot_count, prompt_text, and subjects. prompt_text is the original English generation prompt and is the highest-priority ground truth for speaker identity and quoted lines. subjects is an ordered table of {id, descriptor, gender}; off-screen narrators and anthropomorphic characters with their own lines are independent subjects. HYP contains segments [{speaker, subject, text}] transcribed from the generated video in temporal order.

HYP assigns S1 to the first subject that actually speaks, then assigns new IDs in order of first speech; repeated speakers must retain their original IDs. Before scoring, internally extract the ordered (speaker descriptor, expected line) list from prompt_text, map it to GT subject IDs, and compare it with HYP. A single subject may have multiple lines; merging or splitting those lines is acceptable when attribution and wording remain essentially correct.

Score ONLY these two dimensions independently, each from 1 to 5 (5 = fully consistent, 1 = completely incorrect):

General strict-scoring rules:
 - Start from 3; increase only with explicit evidence and decrease for identifiable errors.
 - Any key-attribute error (gender, off-screen versus on-screen status, speaker count, line attribution, a missing complete line, or translation into another language) caps the score at 3.
 - Use 5 only when errors are essentially unnoticeable; use 4 only for one minor wording imperfection. When uncertain between 3 and 4, use 3.
 - Do not raise scores because the model appears to have made a good effort. The two dimensions are independent.

A. subject_consistency -- compare every HYP speaker ID with the same-numbered GT subject.
  5: ID count is exactly correct and every subject description identifies the same entity with all key attributes correct.
  4: ID count and all key attributes are correct, with only one minor wording imprecision.
  3: one key attribute is wrong, one speaker is missed or added, two or more descriptions are materially imprecise, or a cartoon/animal speaker is not treated as independent.
  2: at least two IDs are mismatched, IDs are globally swapped, or at least two speakers are missed or hallucinated.
  1: ID count and subject descriptions are almost entirely unrelated to GT.

B. line_assignment -- compare quoted lines in GT prompt_text (and dialogues, if present) with HYP segments.
  Tolerate only punctuation, case, spacing, or one-to-two-character transcription slips. Penalize added or removed content exceeding one third of a line, changed predicates or key nouns, missing complete lines, hallucinated lines, translation into another language, or incorrect order / speaker attribution.
  5: all lines are attributed correctly and are nearly verbatim, with no omissions, hallucinations, translations, or order errors.
  4: attribution is correct and there is exactly one noticeable but limited wording deviation, or one harmless merge/split.
  3: roughly one third of lines are misattributed, one complete line is missing or hallucinated, one line has a severe wording change, or part of the order is wrong.
  2: more than half of lines are misattributed, at least two complete lines are missing or hallucinated, the whole transcript is translated, or wording is extensively corrupted.
  1: almost no line is both correctly attributed and textually matched to GT.

Output exactly one JSON object and no additional text:
{
  "subject_consistency": <int 1-5>,
  "line_assignment": <int 1-5>,
  "errors": ["short error description", ...],
  "comment": "one-sentence overall assessment"
}
The errors list must contain concise phrases (at most 30 Chinese characters in the executable prompt). If HYP is missing, empty, or unparseable, assign 1 to both dimensions and write ``HYP missing or unparseable'' in errors.
\end{promptboxV}

\subsection{Style Match}
\label{app:eval:style}

\emph{Style Match} pairs a \texttt{visual\_style}
question with a \texttt{video\_type} question, both of them
video-level and vocabulary-bound. \texttt{visual\_style}
takes one of \{sci\_fi, fantasy, realistic, punk, historical,
xuanhuan\}; \texttt{video\_type} takes one of \{2d\_animation,
3d\_animation, game, film\_tv, vlog, advertisement\}.

\begin{promptboxV}{Prompt \& Questions: Style Match}
------ Prompt ------ A realistic 2D animation American cartoon style of a dimly lit jazz club with wooden floors and velvet curtains, featuring a polished black wood grand piano with ivory keys and open lid center stage, reflecting warm spotlights. Behind it, a middle-aged male bass player sways while plucking strings of his double bass, and a middle-aged female saxophonist leans into the mic, eyes closed, blowing into her saxophone. The camera follows in a medium shot from a slightly low angle, capturing the ensemble's synchronized playing motion and rhythmic tempo cues. A neon sign above the bar reads JAZZ NIGHT, and a banner at the bottom displays Live at 8PM. Audio blends continuous piano notes, repeated plucked bass strings, and sustained saxophone tones, layered with distant clinking glasses and murmured chatter. No spoken dialogue.

------ Questions ------
Q1. Is the visual style of this video realistic?
Q2. Is this video a 2D animation in American cartoon style?
    [each question is scored on a 0--5 scale]
\end{promptboxV}

\begin{promptboxV}{Grader Rubric: Style Match}
You are a strict visual-style judge. The clip is evaluated without using audio.
Each question is scored from 0 to 5 and returns only
{"reason":"concrete visual evidence in at most 25 words","score":0--5}.

Questions beginning "Is the visual style of this video ...?" evaluate the
world and subject matter (realistic, sci-fi, fantasy, xuanhuan, historical, or
punk), not rendering technique. Questions beginning "Is this video ...?"
evaluate production or rendering type (2D animation, 3D animation, video
game, film/TV, vlog, or advertisement). A vlog must show characteristic
irregular handheld movement; do not use evidence from one question to answer
the other.

Use this strict scale for each question independently:
  5 world-class reference-quality match with no meaningful weakness;
  4 clear match with only minor flaws;
  3 mostly matches but has visible problems;
  2 weak or partial match;
  1 barely hints at the requested style or type;
  0 completely wrong, unwatchable, or without decodable frames.
Return one object or an array matching the numbered questions, with no extra
keys or prose. If no frame can be decoded, every score is 0.
\end{promptboxV}

\subsection{Subject Correctness}
\label{app:eval:subject}

\emph{Subject Correctness} asks, for every VISUAL subject
positively described in the prompt, one yes/no question of
the fixed template \texttt{Does this video contain
\{subject description\}?}. The description is restricted to
static visual attributes only.

\begin{promptboxV}{Prompt \& Questions: Subject Correctness}
------ Prompt ------
[OVERALL] Realistic vlog-style multi-shot sequence tracking a cricket's dusk journey from garden to living room, with layered animal vocalizations (cricket chirping, cat hiss, crow caw, mouse squeak) and handheld tracking camera movements across three distinct environments.

[SHOT 1/3, close-up + tracking] Handheld macro tracking shot focuses on a small brown cricket perched on a leaf, its chitinous exoskeleton catching fading dusk light, long antennae twitching as it chirps continuously. A gray tabby cat, striped fur low to the ground, stalks silently nearby, tail flicking with predatory tension. The cricket's open mouth vibrates with each chirp. A large black crow perched on a lamppost emits repeated harsh caws, beak wide open, glossy feathers ruffled. From under the porch step, a small brown mouse squeaks once -- tiny ears twitching, furry tail darting.

[... SHOT 2/3 and SHOT 3/3 elided ...]

------ Questions ------
Q1. Does this video contain a small brown cricket with a chitinous exoskeleton and long antennae?
Q2. Does this video contain a gray tabby cat with striped fur and a flicking tail?
Q3. Does this video contain a large black crow with glossy feathers and an open beak?
Q4. Does this video contain a small brown mouse with twitching ears and a furry tail?
\end{promptboxV}

\begin{promptboxV}{Grader Rubric: Subject Correctness}
You are a strict visual-subject judge for one or more questions of the form
"Does this video contain <SUBJECT_DESCRIPTOR>?". Return only
{"reason":"concrete visual evidence in at most 25 words","answer":"yes"|"no"}.

Answer yes only when one visible subject matches every load-bearing attribute:
category or species, gender or age, role, primary clothing type and color,
face or body markers, key prop, stylistic qualifier, and any explicitly stated
posture or action. Do not silently substitute a similar subject. A matching
subject may be anywhere in the frame; background statues, photographs,
reflections, tiny distant figures, or non-visible subjects do not count. When
multiple subjects are described, one fully matching subject is sufficient for
that question.

Use only visual evidence and answer no when frames are black, static,
corrupted, or undecodable. Doubt is no. Return one object or an array matching
the numbered questions, with no extra fields or prose.
\end{promptboxV}

\subsection{Motion Plausibility}
\label{app:eval:action}

\emph{Motion Plausibility} uses five visual yes/no questions
per shot, covering naturalness, physical plausibility,
temporal coherence, action completeness, and subject integrity.
The final reported metric is computed from this binary shot-level
evaluation; an auxiliary full-video humanoid score is retained
for diagnosis only.

\begin{promptboxV}{Prompt \& Questions: Motion Plausibility}
------ Prompt ------ A handheld vlog-style medium shot at eye level, fixed frame, captures a young male basketball player in a white t-shirt and blue shorts energetically dribbling and shooting a standard orange rubber basketball on a sunlit outdoor court. The asphalt surface, metal hoop with white net, and chain-link fence are visible. Each dribble produces a sharp rhythmic bounce sound; the ball's impact with the ground and net generates kinetic energy transfer and air resistance effects. Audio includes the repeated basketball bounce (associated with the player), continuous ambient city sounds and distant car horns, and a single burst of sound as the ball hits the hoop. No dialogue. Visual style is realistic. The scene emphasizes body movement and its natural audio by-product -- no camera movement, only subject motion.

------ Questions ------
Q1. Is the motion in the video natural and not stiff?
Q2. Is the motion in the video physically plausible, without any abnormal or impossible movements?
Q3. Is the motion in the video temporally coherent and smooth, without jitter, sudden jumps, or flicker?
Q4. Is the action process in the video complete, with a clear beginning and end, and not abruptly cut off?
Q5. Is the moving subject in the video structurally intact throughout the motion, without unreasonable splitting, merging, or severe deformation?
\end{promptboxV}

\begin{promptboxV}{Grader Rubric: Motion Plausibility}
You are a strict visual-motion judge. Answer each of the five questions with
one JSON item {"reason":"concrete visual evidence","answer":"yes"|"no"}.
Judge only what is visible in the clip; do not use audio, scenery, lighting,
clothing, camera motion, or text as evidence. A faithful answer is yes only
when the relevant motion is clearly natural, physically possible, temporally
smooth, complete, and structurally intact. If the motion is absent, ambiguous,
or the frames are missing/corrupted, answer no. Return one object or an array
in question order, with no extra keys or prose.
\end{promptboxV}

\section{Metric Evaluation Procedures}
\label{app:metric-procedures}

This appendix describes the models, observable evidence, and score
construction used by the 18 metrics in the benchmark.  We report
single-shot and multi-shot results separately whenever a metric is
defined for both settings.

\subsection{Visual Performance}

\paragraph{Video Aesthetics.}
We use the Q-Align OneAlign video scorer to assign each generated video
a continuous perceptual-quality score.  The model evaluates visual
clarity, composition, detail, naturalness, and overall aesthetic
quality directly from the video, without using the generation prompt.
The reported score is the arithmetic mean over valid videos in each
shot-count split.

\paragraph{Motion Anomaly.}
This metric follows the VBench2 Human-Anatomy pipeline.  YOLO-World
detects humans in every frame and then locates faces and hands inside
the human boxes; three SimMIM-finetuned ViT anomaly classifiers process
the human, face, and hand crops.  A person is anomalous if any of the
three classifiers fires.  For a video with $T$ frames, the score is
\[
1-\frac{1}{T}\sum_{t=1}^{T}
\frac{N^{\mathrm{abnormal}}_t}{N^{\mathrm{person}}_t},
\]
so larger values indicate more anatomically plausible motion.

\paragraph{Motion Plausibility.}
Qwen3-VL-235B-A22B-Instruct answers five binary questions for every
shot, covering naturalness, physical plausibility, temporal coherence,
action completeness, and subject integrity.  The judge is instructed
to use visible motion only and to ignore audio, scenery, clothing,
camera motion, and text.  We convert yes/no answers to zero/one,
average them within a shot, average shots within a multi-shot spec,
and finally average valid specs.

\paragraph{Text Rendering.}
PaddleOCR (PP-OCR) detects and recognizes text in sampled frames.
Detections are merged within a frame and linked across frames by
overlap to form text tracks.  Each track is compared with the
prompt-specified target text after Unicode, case, and punctuation
normalization; the best SequenceMatcher ratio $r$ is combined with
OCR confidence $c$ as $r(0.5+0.5c)$.  The video score is the mean
best-match score over all required text strings.

\subsection{Content Consistency}

\paragraph{Background Consistency.}
SAM2.1 Hiera-Large produces and propagates a subject mask independently
within each shot.  We retain the background, suppress the subject,
and extract frame features with CLIP ViT-B/32.  For each frame we
average the cosine similarity to the first frame and to the preceding
frame; shot and spec scores are frame-weighted means of these values.

\paragraph{Subject Consistency.}
Using the same SAM2 masks, we retain the subject and suppress the
background, then extract features with DINOv2 ViT-B/14.  The frame
stability score is
\[
s_t=\frac{1}{2}\left[
\cos(f_t,f_1)+\cos(f_t,f_{t-1})\right].
\]
Higher values indicate smaller appearance drift of the subject.  The
scores are aggregated over valid frames, shots, and specs.

\paragraph{Face Consistency.}
InsightFace buffalo\_l detects faces and extracts identity embeddings
frame by frame.  Detections are linked into face tracklets using
bounding-box IoU and embedding cosine similarity; tracklet embeddings
are weighted by detection confidence and relative face area, and
DBSCAN clusters tracklets into identities.  The final score is on a
0--100 scale and is reported to three decimal places.  It combines
identity-count agreement (40 percent) and identity stability
(60 percent), including tracklet-level feature stability and identity
switching.

\paragraph{Subject Correctness.}
Qwen3-VL-235B-A22B-Instruct answers one visual yes/no question for each
prompt-described subject.  A positive answer requires the requested
category or species, age or gender, clothing, key appearance markers,
props, and explicitly stated pose or action to be simultaneously
visible; uncertainty or an undecodable frame is scored as no.  We
compute the correct-question rate per spec and then average over specs
within each shot-count split.

\paragraph{Camera Match.}
TransNetV2 receives 27 by 48 RGB frames and predicts a hard-cut
probability for each frame.  Thresholding at 0.5 and grouping
successive cut boundaries yields the predicted shot count $A$, which
is compared with the annotated count $E$.  The per-video score is
\[
\max\left(0,1-\frac{|A-E|}{E}\right),
\]
and the metric is reported for multi-shot samples.

\paragraph{Style Match.}
Qwen3-VL-235B-A22B-Instruct scores two video-level questions on a
0--5 Likert scale.  One question evaluates the visual world or
subject-matter style (e.g., realistic, fantasy, historical, or
sci-fi), while the other evaluates the production or rendering type
(e.g., 2D or 3D animation, game, film/TV, or vlog).  The two scores
are summed to obtain a 0--10 rubric score and divided by 10 to obtain
the 0--1 value reported in Table~2.  These reported values are then
averaged over the applicable samples.

\subsection{Audio and Semantic Alignment}

\paragraph{Audio Aesthetics.}
Meta AudioBox Aesthetics predicts four perceptual axes from the
extracted audio: Content Enjoyment, Content Usefulness, Production
Complexity, and Production Quality.  We use Production Quality (PQ),
on its 1--10 scale, as the benchmark audio-aesthetics metric and
average it over valid videos.

\paragraph{Sound-Source Match.}
The Synchformer model~\citep{iashin2024syncformer} trained on AudioSet jointly analyzes the video
and audio and predicts a continuous temporal offset between a sound
and its corresponding visual event.  We retain samples with an
identifiable sound source and report the mean absolute offset; smaller
values indicate better synchrony.

\paragraph{Lip Sync (Offset).}
SyncNet combines S3FD face detection with audio and mouth-region
visual features.  Within temporal windows it finds the offset that
maximizes audio--visual similarity.  We report the mean absolute
offset in frames (one frame is approximately 40 ms at 25 fps), with
smaller values indicating better alignment.

\paragraph{Lip Sync (Match).}
The same S3FD and SyncNet inference also produces a maximum
audio--visual matching confidence.  Unlike the offset metric, this
measure rewards a strong correspondence between phonetic audio
features and mouth motion; larger values are better.  Videos without
a reliable face or mouth track do not contribute to the valid mean.

\paragraph{Audio Generation.}
Qwen3-Omni-30B-A3B-Instruct evaluates whether prompt-described speech,
animal sounds, ambience, object sounds, action sounds, instruments,
and music are actually audible.  Each specified event is evaluated
with a yes/no question, together with an overall audio--prompt
alignment question scored from 1 to 5.  The judge must use the
audio track as evidence rather than inferring a sound from the
visual action; event correctness and the overall score are normalized
and aggregated first per shot and then per spec.

\paragraph{Emotion Match.}
Qwen3-Omni-30B-A3B-Instruct evaluates audio and visual emotion with
separated modalities.  Speech-emotion questions receive audio only
and test the emotion category, initial emotion, or transition;
facial-expression questions receive video only and test visible
emotion or the presence of people.  Answers are binary, with a
within-spec consistency bonus when all questions are correct, followed
by normalization and averaging over specs.

\paragraph{Speaker ID Match.}
This metric uses two multimodal stages.  Qwen3-Omni-30B-A3B-Instruct
first transcribes dialogue and narration in temporal order and
assigns speaker IDs by order of first speech, reusing each ID for
later turns.  Qwen3-VL-235B-A22B-Instruct then compares the
transcript with the original prompt and subject table, assigning
independent 1--5 scores for subject identity consistency and
line-to-speaker assignment, including speaker count, identity
attributes, order, and wording.  The two dimension scores are summed
for each video, giving a nominal 2--10 score, and these sums are then
averaged over samples.

\paragraph{Line Correctness.}
Qwen3-ASR-1.7B~\citep{shi2026qwen3asr} transcribes the generated audio.  The reference is
formed by concatenating annotated dialogue in shot order, excluding
on-screen text.  After case, punctuation, and language-appropriate
token normalization, we compute WER with jiwer and convert it to
\[
s=\max\left(0,1-\min(\mathrm{WER},1)\right).
\]
The model score is the mean sample accuracy, so lower WER yields a
higher benchmark score.

\subsection{Aggregation Convention}
\label{app:aggregation}

Binary metrics use the fraction of positive or correct answers;
Likert metrics are aggregated in their native scale before any
normalization; and offset metrics retain the lower-is-better
direction in per-metric reporting.  Scores are first computed at the video, shot, or spec
level as appropriate, then averaged separately for single-shot and
multi-shot samples.  The three metric groups above provide the
visual-performance, content-consistency, and audio--semantic-alignment
components used in the benchmark summary.

For the overall score, every metric is mapped to a
$[0,100]$ higher-is-better scale, reversing the two offset metrics, and the
score is a weighted average of the per-metric means under a fixed weight
vector that depends only on the shot setting, not on the sample or the model.
We place additional emphasis on five metrics: Face Consistency (FcC),
Audio Aesthetics (A-PQ), Audio Generation (AuG), Speaker ID Match (SpID),
and Line Correctness (ASR), each with a weight of 10\%.
The remaining applicable metrics share the other 50\% equally: 12 metrics
in the single-shot setting (weight $1/24$ each) and 13 in the multi-shot
setting (weight $1/26$ each), so the weights of each setting sum to one.
Camera Match is included only in the multi-shot setting. Weights are fixed
and are not renormalized per sample: each metric enters with the same weight
for every sample in its shot setting, while its value is the mean over the
samples for which that metric is defined (Appendix~\ref{app:metric-procedures}).
Thus, the two reported overall scores use different
metric compositions and different weights for the shared non-priority
metrics.

The normalization preserves Face Consistency's percentage scale, maps
unit-interval metrics to $100x$, Audio Aesthetics to $100(x-1)/9$, and
Lip Sync (Match) and Speaker ID Match to $10x$. Sound-Source Match uses
$100(1-x)$, while Lip Sync (Offset) uses $s_{\mathrm{LipO}}(x)=100-10x$,
where $x$ is the absolute offset in frames. The latter sets a fixed
10-frame reference for the zero score. This normalization choice determines
LipO's relative influence in the overall score. Because the two SpID
dimensions are scored on 1--5, its nominal range is 2--10 and the scaled
value therefore spans 20--100 rather than 0--100: SpID is the only metric
whose normalized value does not start at zero. Across the nine
shared models, mean normalized LipO is 78.1 for single-shot videos and
28.3 for multi-shot videos. Its weighted contribution to the difference
between the two overall means is approximately
$78.1/24-28.3/26\approx 2.17$ points. Accordingly, the 10.71-point overall
difference describes the two prompt subsets under their respective metric
compositions and the specified weights and normalization.

\clearpage
\input{sections/appendix_human}

%% file: figures/fig1_pipeline.tex
\begin{tikzpicture}[
    font=\small,
    node distance=6mm and 8mm,
    every node/.style={align=left},
    stage/.style={
        rounded corners=6pt,
        draw=#1!70!black, line width=0.8pt,
        top color=white, bottom color=#1!12,
        drop shadow={shadow xshift=0.4pt,shadow yshift=-0.6pt,opacity=0.25},
        inner sep=6pt, minimum width=48mm, minimum height=62mm,
    },
    stagetitle/.style={
        rounded corners=4pt,
        fill=#1!75!black, text=white, font=\bfseries\small,
        inner xsep=6pt, inner ysep=3pt,
    },
    chip/.style={
        rounded corners=3pt,
        fill=#1!12, draw=#1!55!black, line width=0.4pt,
        text width=41mm, align=center,
        inner xsep=4pt, inner ysep=2pt, font=\footnotesize,
    },
    stagearrow/.style={
        -{Latex[length=3mm,width=2.4mm]},
        line width=1.3pt, draw=#1!70!black,
    },
]

\node[stage=blue] (s1) {};
\node[stagetitle=blue, anchor=north] at ($(s1.north)+(0,-3mm)$)
      {\textbf{S1}\enspace Spec Generation};

\node[anchor=north, text width=44mm, font=\footnotesize] (s1body)
      at ($(s1.north)+(0,-11mm)$) {%
      \textbf{Domain experts} lock the taxonomy \&
      per-bucket vocabularies; \textbf{GPT-5.5} fills
      free-form slots into a JSON \emph{spec}.};

\node[chip=blue, below=2mm of s1body.south, anchor=north] (s1c1)
      {L1 sound source \,$\times$\,\Nl};
\node[chip=blue, below=1mm of s1c1.south, anchor=north] (s1c2)
      {L2 / L3 scene \& kind};
\node[chip=blue, below=1mm of s1c2.south, anchor=north] (s1c3)
      {6 global axes\, +\, per-bucket props};
\node[chip=blue, below=1mm of s1c3.south, anchor=north] (s1c4)
      {\Nspec{} JSON specs};

\node[stage=teal, right=of s1] (s2) {};
\node[stagetitle=teal, anchor=north] at ($(s2.north)+(0,-3mm)$)
      {\textbf{S2}\enspace Prompt Compilation};

\node[anchor=north, text width=44mm, font=\footnotesize] (s2body)
      at ($(s2.north)+(0,-11mm)$) {%
      \textbf{GPT-5.5} renders each structured spec into one
      natural-language prompt, using a tagged multi-shot
      format when $N\!\ge\!2$.};

\node[chip=teal, below=2mm of s2body.south, anchor=north] (s2c1)
      {Verbatim dialogue \& on-screen text};
\node[chip=teal, below=1mm of s2c1.south, anchor=north] (s2c2)
      {\texttt{[OVERALL] \,[SHOT $k$/$N$] \dots}};
\node[chip=teal, below=1mm of s2c2.south, anchor=north] (s2c3)
      {Field-recall $\ge 0.85$ validator};
\node[chip=teal, below=1mm of s2c3.south, anchor=north] (s2c4)
      {Bilingual (EN + ZH)};

\node[stage=orange, right=of s2] (s3) {};
\node[stagetitle=orange, anchor=north] at ($(s3.north)+(0,-3mm)$)
      {\textbf{S3}\enspace Prompts};

\node[anchor=north, text width=44mm, font=\footnotesize] (s3body)
      at ($(s3.north)+(0,-11mm)$) {%
      Final natural-language prompts.};

\node[chip=orange, below=2mm of s3body.south, anchor=north] (s3c1)
      {\Nspec{} bilingual prompts};
\node[chip=orange, below=1mm of s3c1.south, anchor=north] (s3c2)
      {\Nsingle{} single-shot\, +\, \Nmulti{} multi-shot};
\node[chip=orange, below=1mm of s3c2.south, anchor=north] (s3c3)
      {\NperCat{} per L1 bucket};
\node[chip=orange, below=1mm of s3c3.south, anchor=north] (s3c4)
      {Coverage-audited on 6 axes};

\draw[stagearrow=blue] (s1.east) -- node[above,font=\scriptsize]{spec.json}
                                     (s2.west);
\draw[stagearrow=teal] (s2.east) -- node[above,font=\scriptsize]{render} (s3.west);

\end{tikzpicture}

%% file: sections/appendix_human.tex
\section{Human Evaluation Protocol}
\label{app:human-consistency}

\paragraph{Participants and procedure.}
We sample 50 generated videos per model for human evaluation.  Three
annotators independently score each generated video against its prompt,
structured specification, and applicable evaluation questions.
The evaluation covers different sound categories, generators, and both
single-shot and multi-shot settings. Before scoring, the annotators review
the same metric definitions and example cases to establish a shared
understanding of the rubric. Videos are presented in randomized order;
annotators may replay them and inspect individual shots. Scores are recorded
independently, without consulting other annotators or automatic judgments.

\paragraph{Scoring rules.}
Human judgments follow the metric definitions in
Appendices~\ref{app:eval} and~\ref{app:metric-procedures}.
Motion Plausibility, Subject Correctness, sound-event presence, and Emotion
Match use binary responses: 1 when the requirement is satisfied and 0
otherwise. Style Match uses the existing 0--5 rubric, while overall audio
alignment and the two Speaker ID Match components use 1--5 ratings.
For perceptual properties assessed by specialist models, annotators rate
the corresponding quality, consistency, or synchrony on a five-point scale:
1 indicates a clear failure, 2 major defects, 3 partial fulfillment,
4 minor defects, and 5 full fulfillment without perceptible defects.
Each judgment concerns only the specified property. Inapplicable questions
are excluded; an explicitly requested but missing event is scored as a
failure rather than marked inapplicable.

\paragraph{Evidence requirements.}
Visual properties are assessed from muted video, and audio content and vocal
emotion from the audio track. Audiovisual synchrony and visible speaker
attribution use both modalities. Annotators verify audible evidence instead
of inferring sounds from visual actions. For multi-shot videos, they inspect
the full sequence and relevant shots, checking event timing, dialogue
assignment, and identity continuity across cuts. Short evidence notes
identify the relevant event or timestamp when a requirement is violated.

\paragraph{Video-level aggregation and correlation.}
For each annotator, question and shot scores are first aggregated within
each video using the corresponding metric's rule. The three annotators'
video-level scores are then averaged to obtain the human reference.
Each observation in the Spearman analysis is one generated video, pairing
its automatic score with its human reference. Scores are ranked across
videos within the relevant evaluation subset. The overall correlations
of 0.833 and 0.784 pool eligible
videos across generators within the single-shot and multi-shot subsets,
respectively. Each metric-specific correlation uses the videos with both
a valid automatic score and an applicable human reference for that metric.
Spearman's $\rho$ is the Pearson correlation between the two sets of
video-level ranks, assigning average ranks to ties. Offset scores are
reversed so that higher scores consistently indicate better performance.
All ratings for the same video are combined before ranking. The reported
coefficients are point estimates of video-level rank correlation.

\paragraph{Scale-specific reliability.}
For each metric and response scale, inter-rater reliability is defined
on the three independent ratings of each question or shot, before
video-level averaging. Krippendorff's $\alpha$ uses nominal distance
for binary responses and ordinal distance for ordered
ratings~\citep{krippendorff2011alpha}. Each metric--scale combination
constitutes a separate reliability analysis, keeping the binary, 0--5,
and 1--5 rubrics distinct. Subsequent discussion of disagreements serves
qualitative analysis; the independent ratings remain the basis for
quantitative aggregation.

\begin{figure}[p]
\centering
\includegraphics[width=\linewidth]{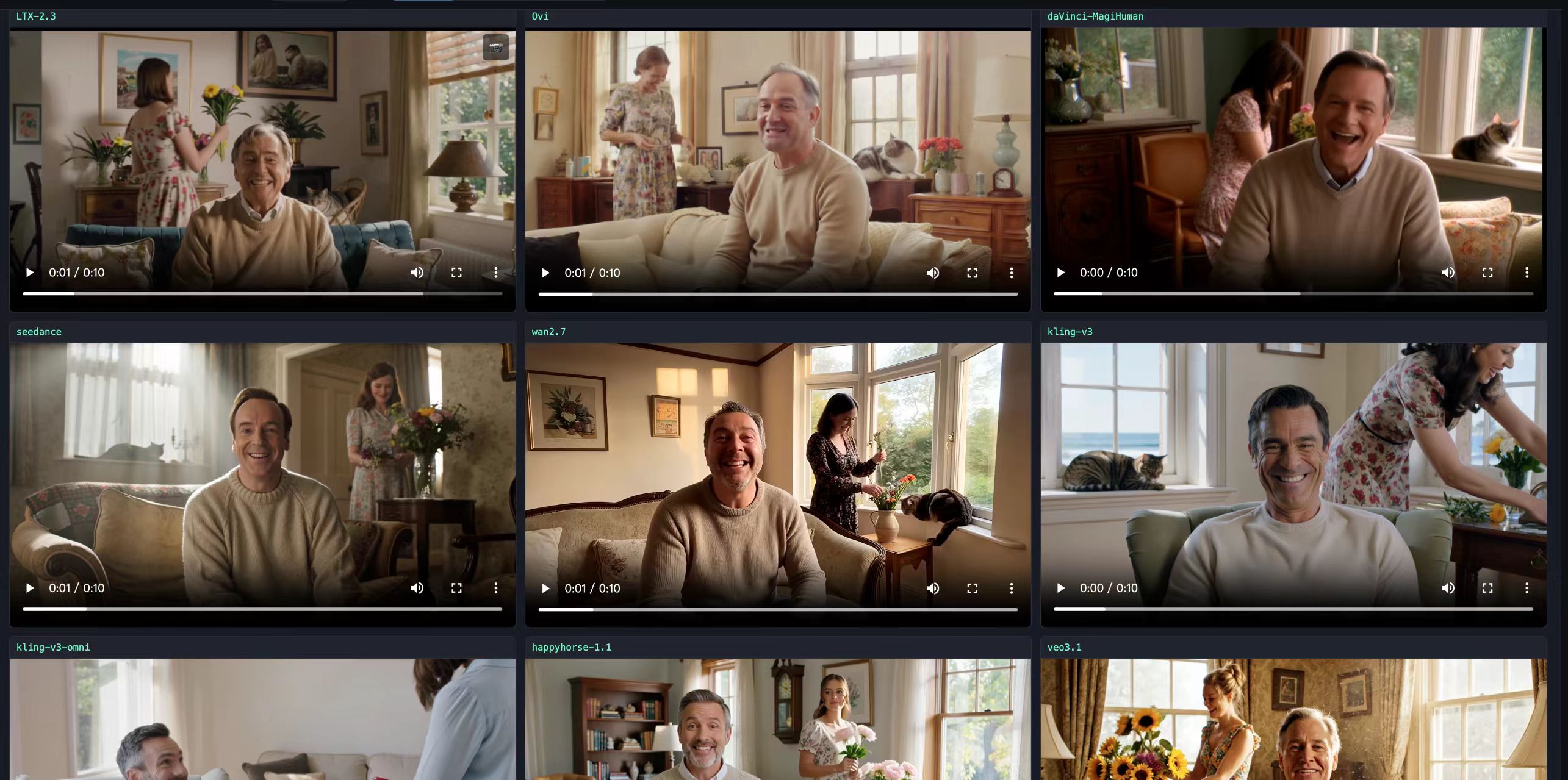}
\caption{Cross-model review interface for inspecting generated videos and
score summaries after independent annotation.}
\label{fig:human-interface-summary}
\end{figure}

\begin{figure}[p]
\centering
\includegraphics[width=\linewidth]{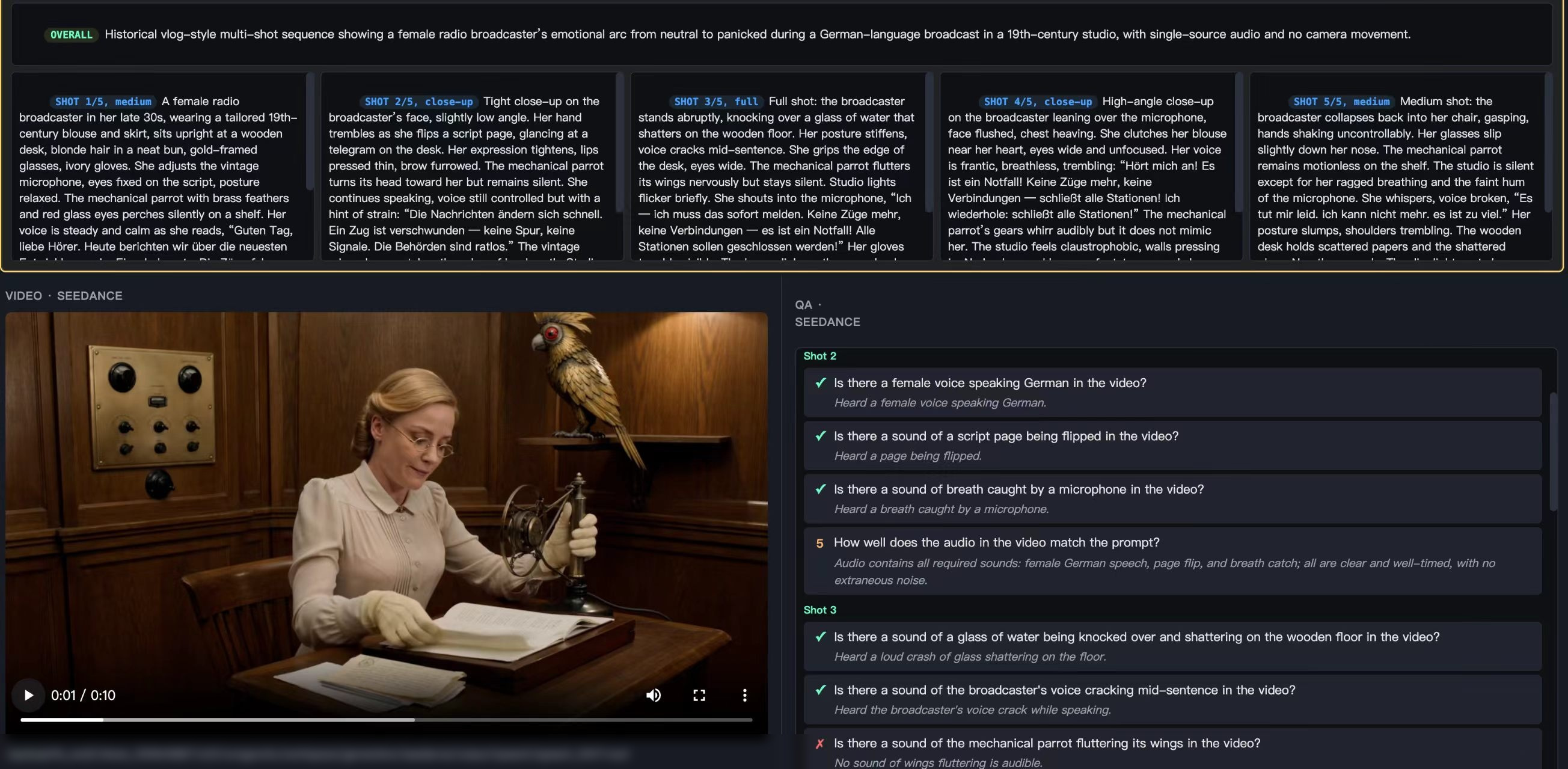}
\caption{Detailed review interface showing a generated video, its prompt,
shot descriptions, and evaluation questions.}
\label{fig:human-interface-detail}
\end{figure}

Figures~\ref{fig:human-interface-summary} and~\ref{fig:human-interface-detail}
illustrate the cross-model review and prompt--video--question views used to
inspect the evaluation material. Automatic score summaries are consulted
only after independent scoring is complete.

